\documentstyle[12pt]{article}

\catcode`\@=11
\def\marginnote#1{}
\newcount\hour
\newcount\minute
\newtoks\amorpm
\hour=\time\divide\hour by60
\minute=\time{\multiply\hour by60 \global\advance\minute
by-\hour}\edef\standardtime{{\ifnum\hour<12
\global\amorpm={am}%
        \else\global\amorpm={pm}\advance\hour by-12 \fi
        \ifnum\hour=0 \hour=12 \fi
        \number\hour:\ifnum\minute<10
0\fi\number\minute\the\amorpm}}
\edef\militarytime{\number\hour:\ifnum\minute<10
0\fi\number\minute}

\def\draftlabel#1{{\@bsphack\if@filesw {\let\thepage\relax
   \xdef\@gtempa{\write\@auxout{\string
      \newlabel{#1}{{\@currentlabel}{\thepage}}}}}\@gtempa
   \if@nobreak \ifvmode\nobreak\fi\fi\fi\@esphack}
        \gdef\@eqnlabel{#1}}
\def\@eqnlabel{}
\def\@vacuum{}
\def\draftmarginnote#1{\marginpar{\raggedright\scriptsize\tt#1}}
\def\draft{\oddsidemargin -.5truein
        \def\@oddfoot{\sl preliminary draft \hfil
        \rm\thepage\hfil\sl\today\quad\militarytime}
        \let\@evenfoot\@oddfoot \overfullrule 3pt
        \let\label=\draftlabel
        \let\marginnote=\draftmarginnote

\def\@eqnnum{(\theequation)\rlap{\kern\marginparsep\tt\@eqnlabel}%
\global\let\@eqnlabel\@vacuum}  }

\def\numberbysection{\@addtoreset{equation}{section}
        \def\theequation{\thesection.\arabic{equation}}}

\def\underline#1{\relax\ifmmode\@@underline#1\else
 $\@@underline{\hbox{#1}}$\relax\fi}

\catcode`@=12
\relax

\numberbysection

\advance \topmargin by -\headheight
\advance \topmargin by -\headsep

\evensidemargin \oddsidemargin
\def\rf#1{(\ref{#1})}
\def\lab#1{\label{#1}}
\def\nonu{\nonumber}
\def\br{\begin{eqnarray}}
\def\er{\end{eqnarray}}
\def\be{\begin{equation}}
\def\ee{\end{equation}}

\def\llbrack{\lbrack}
\def\rrbrack{\rbrack}

\def\({\left(}
\def\){\right)}

\newcommand{\ct}[1]{\cite{#1}}
\newcommand{\bi}[1]{\bibitem{#1}}

\relax

\def\tp0{\Theta_{+}^{(0)}}
\def\tm0{\Theta_{-}^{(0)}}

\def\f#1#2#3 {f^{#1#2}_{#3}}

\def\win1{{\sf w_{1+\infty}}}

\def\Win1{{\sf W_{1+\infty}}}

\def\rlx{\relax\leavevmode}
\def\inbar{\vrule height1.5ex width.4pt depth0pt}
\def\IZ{\rlx\hbox{\sf Z\kern-.4em Z}}
\def\IR{\rlx\hbox{\rm I\kern-.18em R}}
\def\IC{\rlx\hbox{\,$\inbar\kern-.3em{\rm C}$}}
\def\IN{\rlx\hbox{\rm I\kern-.18em N}}
\def\IO{\rlx\hbox{\,$\inbar\kern-.3em{\rm O}$}}
\def\IP{\rlx\hbox{\rm I\kern-.18em P}}
\def\IQ{\rlx\hbox{\,$\inbar\kern-.3em{\rm Q}$}}
\def\IF{\rlx\hbox{\rm I\kern-.18em F}}
\def\IG{\rlx\hbox{\,$\inbar\kern-.3em{\rm G}$}}
\def\IH{\rlx\hbox{\rm I\kern-.18em H}}
\def\II{\rlx\hbox{\rm I\kern-.18em I}}
\def\IK{\rlx\hbox{\rm I\kern-.18em K}}
\def\IL{\rlx\hbox{\rm I\kern-.18em L}}
\def\one{\hbox{{1}\kern-.25em\hbox{l}}}
\def\0#1{\relax\ifmmode\mathaccent"7017{#1}%
B        \else\accent23#1\relax\fi}

\def\PRL#1#2#3{{\sl Phys. Rev. Lett.} {\bf#1} (#2) #3}
\def\NPB#1#2#3{{\sl Nucl. Phys.} {\bf B#1} (#2) #3}

\def\CMP#1#2#3{{\sl Commun. Math. Phys.} {\bf #1} (#2) #3}

\def\PRE#1#2#3{{\sl Phys. Rev.} {\bf E#1} (#2) #3}

\def\JMP#1#2#3{{\sl J. Math. Phys.} {\bf #1} (#2) #3}

\def\InvM#1#2#3{{\sl Invent. Math.} {\bf #1} (#2) #3}

\def\IJMPA#1#2#3{{\sl Int. J. Mod. Phys.} {\bf A#1} (#2) #3}
\def\AdM#1#2#3{{\sl Advances in Math.} {\bf #1} (#2) #3}

\def\JETPL#1#2#3{{\sl  Sov. Phys. JETP Lett.} {\bf #1} (#2) #3}

\def\PHSD#1#2#3{{\sl Physica} {\bf D#1} (#2) #3}

\def\JPSJ#1#2#3{{\sl J. Phys. Soc. Japan} {\bf #1} (#2) #3}

\begin{document}
\begin{titlepage}
\vspace*{-1cm}

\vspace{.2in}
\begin{center}
{\large\bf Vector NLS hierarchy solitons revisited: dressing transformation and tau function approach}
\end{center}

\vspace{1in}

\begin{center}
Harold Blas

\vspace{.5 cm}
\small

\par \vskip .1in \noindent
Instituto de F\'\i sica Te\'orica - IFT/UNESP\\
Rua Pamplona 145\\
01405-900  S\~ao Paulo-SP, BRAZIL

\normalsize
\end{center}

\vspace{1.5in}

\begin{abstract}
We discuss some algebraic aspects of the integrable vector non-linear Schr\"{o}dinger hierarchies (GNLS$_{r}$). These are hierarchies of zero-curvature equations constructed from affine Kac-Moody algebras $\hat{sl}_{r+1}$. Using the dressing transformation method and the tau-function formalism, we construct the N-soliton solutions of the GNLS$_{r}$ systems. The explicit matrix elements in the case of GNLS$_{1}$ are computed using level one vertex operator representations. 
\end{abstract}
\end{titlepage}
\section{Introduction}
It is well known that the $1+1$-dimensional non-linear Schr\"{o}dinger ({\bf NLS}) equation is integrable and possesses exact soliton solutions \ct{zakharov}. It has been known that many soliton equations in $1+1$ dimensions have integrable matrix generalizations, or more generally, integrable multi-component generalizations. The most well-known example for the multi-component case is the vector {\bf NLS} equation, first studied by Manakov \ct{manakov}. These type of equations and their higher-order generalizations find applications in non-linear optics (for a complete review of the most important references in the field see \ct{nls}). The multi-soliton type solutions of these hierarchies can be obtained using diverse methods. For example, in \ct{zen} the vector {\bf NLS} equation has been studied in the framework of the inverse scattering method, the bright and dark multi-soliton solutions and their collisions have been studied.     

One of the most fascinating applications of the Kac-Moody theory and its affine Lie algebras and their relevant groups is to exhibit hidden symmetries of soliton equations. According  to the approach of \ct{ferreira} a common feature of integrable hierarchies presenting soliton solutions is the existence of some special ``vacuum solutions'' such that the Lax operators evaluated on them lie in some abelian subalgebra of the associated Kac-Moody algebra. The soliton type solutions are constructed out of those ``vacuum solutions'' through the so called dressing transformation procedure. These developments lead to a quite general definition of  tau functions associated to the hierarchies, in terms of the so called ``integrable highest weight representations'' of the relevant Kac-Moody algebra.
 
In this paper we obtain the multi-soliton solutions of the vector {\bf NLS} equation  using the dressing transformation method. We believe that the group theoretical point of view of finding the analitical results for the general case of $N$-soliton interactions could facilitate the study of their properties; for example, the asymptotic behaviour of trains of $N$ solitonlike pulses with approximately equal amplitudes and velocities, as studied in \ct{kaup1}. The second point we shoul highlight relies upon the possible relevance of the {\bf NLS} tau functions to its higher-order generalizations. We believe that the tau functions of the higher-order {\bf NLS} generalization are related somehow to the basic tau functions of the usual (vector) {\bf NLS} equations (this fact is observed for example in the case of the coupled {\bf NLS}$+${\bf DNLS} system \ct{blas, liu}, in the second Ref. Hirota's method has been used).

The plan is as follows. In section $2$ we review the theory of the dressing transformations and the definition of the tau-function vectors. In section 3 we present the construction of the vector {\bf NLS} equations (GNLS$_{r}$) associated to the homogeneous gradation of the Kac-Moody algebra $\hat{sl}_{r+1}$, their relevant tau functions are defined and the construction of multiple-soliton solutions uotlined. In section 4 we present a detailed study of the GNLS$_{1}$ case; the first conserved charges are constructed in the context of this formalism and the explicit form of the $N$-soliton solutions are presented. Finally, for the sake of completeness, we have included Kac-Moody algebra notations and conventions, as well as, its well known theory of integrable highest weight representations, and level one homogeneous vertex operator representations (see appendices \ref{appa}, \ref{appb}, \ref{appc}). Some of the details regarding the matrix elements appear in the appendices \ref{appd}, \ref{appe} and \ref{appf}.

\section{Dressing Transformations}

Consider a non linear system which can be formulated in terms of a system of
first-order differential equations 
\begin{equation}
{\cal L}_N{\bf \Psi =}0,  \lab{dr0}
\end{equation}
where ${\cal L}_N$ are Lax operators of the form 
\begin{equation}
\qquad {\cal L}_N=\frac \partial {\partial _{t_N}}-A_N  \lab{dr1}
\end{equation}
and the variables $t_N$ are the various ``times'' of the hierarchy.

Then, the equations of the hierarchy are equivalent to the integrability or
zero curvature conditions of \rf{dr0}, 
\begin{equation}
\llbrack {\cal L}_N\, ,\, {\cal L}_M \rrbrack = 0.  \lab{dr2}
\end{equation}

Therefore the Lax operators are of the form of a pure gauge

\begin{equation}
A_N=\frac{\partial {\bf \Psi }}{\partial _{t_N}}{\bf \Psi }^{-1}.  \lab{dr3}
\end{equation}

The type of integrable hierarchy considered here is based on a Kac-Moody
algebra $\widehat{g}$ furnished with an integer gradation labelled by a
vector $s=(s_o,s_1,...,s_r)$ of $r+1$ non-negative co-prime integers such
that

\br
\widehat{g}=\bigoplus_{i\in \IZ}\widehat{g}_i(s) \qquad \mbox{and}\qquad 
\llbrack \widehat{g}_i(s)\, ,\, \widehat{g}_j(s)\rrbrack \subseteq \widehat{g}_{i+j}(s).
\lab{dr4}
\er

The connections we consider are of the form

\begin{equation}
A_N=\sum_{i=0}^lA_{N,i}\qquad \mbox{where}\qquad A_{N,i}\in \widehat{g}_i(s).
\lab{dr5}
\end{equation}

The ``dressing transformations'' are non local gauge transformations on $%
A_N $ which maintain their form and gradation \ct{ferreira, babelon}.
Each of these gauge transformations is made with the help of two group
elements $\Theta _{+}$ and $\Theta _{-}$ , such that 
\begin{equation}
A_N\rightarrow A_N^h\equiv \Theta _{\pm }A_N\Theta _{\pm }^{-1}+\partial
_N\Theta _{\pm }\Theta _{\pm }^{-1}  \lab{dr6}
\end{equation}
or 
\begin{equation}
A_N^h=\frac{\partial (\Theta _{\pm }{\bf \Psi )}}{\partial _{t_N}}(\Theta
_{\pm }{\bf \Psi )}^{-1}  \lab{dr7}
\end{equation}

We have a residual gauge transformation in (\ref{dr7})

\begin{equation}
{\bf \Psi \rightarrow \Psi }h,  \lab{dr8}
\end{equation}
where $h$ is a constant group element. Therefore we can impose
\begin{equation}
\Theta _{+}{\bf \Psi }=\Theta _{-}{\bf \Psi }h  \label{dr9}
\end{equation}
or equivalently
\begin{equation}
\Theta _{-}^{-1}\Theta _{+}={\bf \Psi }h{\bf \Psi }^{-1}  \label{dr10}
\end{equation}

$\Theta _{-}\Psi $ \,\, defines a new solution
\begin{equation}
{\bf \Psi }^h=\Theta _{-}{\bf \Psi }  \label{dr11}
\end{equation}

We admit a Gauss decomposition with respect to the gradation

\begin{equation}
{\bf \Psi }h{\bf \Psi }^{-1}=\left( {\bf \Psi }h{\bf \Psi }^{-1}\right)
_{-}\left( {\bf \Psi }h{\bf \Psi }^{-1}\right) _0\left( {\bf \Psi }h{\bf %
\Psi }^{-1}\right) _{+}  \label{dr12}
\end{equation}
We choose (see (\ref{dr10}))

\begin{equation}
\Theta _{-}=(\left( {\bf \Psi }h{\bf \Psi }^{-1}\right) _{-})^{-1}
\label{dr13}
\end{equation}
and therefore (\ref{dr11}) can be written as 
\begin{equation}
{\bf \Psi }^h=(\left( {\bf \Psi }h{\bf \Psi }^{-1}\right) _{-})^{-1}{\bf %
\Psi =}\Theta _{+}{\bf \Psi }h^{-1}=\left( {\bf \Psi }h{\bf \Psi }%
^{-1}\right) _0\left( {\bf \Psi }h{\bf \Psi }^{-1}\right) _{+}{\bf \Psi }%
h^{-1}  \label{dr14}
\end{equation}
where we used (\ref{dr9}) and (\ref{dr10}). $\Psi ^h$ in (\ref{dr14}) is also a
solution of the linear problem (\ref{dr1}).

We shall consider solutions which belong to the orbits of the vacuum
solutions.

Define
\br
\Theta _{-}^{-1}=\left( {\bf \Psi }^{\left( vac\right) }h{\bf \Psi }^{\left(
vac\right) -1}\right) _{-},\quad B^{-1}=\left( {\bf \Psi }^{\left(
vac\right) }h{\bf \Psi }^{\left( vac\right) -1}\right) _0 
\er
\begin{equation}
N=\left( {\bf \Psi }^{\left( vac\right) }h{\bf \Psi }^{\left( vac\right)
-1}\right) _{+}\quad \mbox{and}\quad \Omega =B^{-1}N.  \label{dr15}
\end{equation}

Therefore the dressing transformations generated by $h$ becomes 
\begin{equation}
{\bf \Psi }^{\left( vac\right) }\rightarrow {\bf \Psi }^h=\Theta _{-}{\bf %
\Psi }^{\left( vac\right) }=\Omega {\bf \Psi }^{\left( vac\right) }h^{-1}.
\label{dr16}
\end{equation}

Denote by $\mid \lambda _i>$ the state of highest weight of a fundamental
representation such that $s_i\neq 0$.

Define the tau-function vector \ct{ferreira}
\begin{eqnarray}
\tau _i(t^{\pm }) &=&\left( {\bf \Psi }^{\left( vac\right) }h{\bf \Psi }%
^{\left( vac\right) -1}\right) \left| \lambda _i\right\rangle  \nonumber \\
&=&\Theta _{-}^{-1}B^{-1}\left| \lambda _i\right\rangle ,\qquad \qquad
i=0,1,...,r;\qquad s_i\neq 0.  \label{dr17}
\end{eqnarray}

Note that $N\mid \lambda _i>=\mid \lambda _i>$ and 
\begin{equation}
\tau _i^{\left( 0\right) }(t^{\pm })=B^{-1}\left| \lambda _i\right\rangle
=\left| \lambda _i\right\rangle \widehat{\tau }_i^{\left( 0\right) }(t^{\pm
}),  \label{dr18}
\end{equation}
{\bf \ }
since $\mid \lambda _i>$ is an eigenstate of the sublagebra $\widehat{g}
_o(s) $. Then
\begin{equation}
\widehat{\tau }_i^{\left( 0\right) }(t^{\pm })=\left\langle \lambda
_o\right| \left[ {\bf \Psi }^{\left( vac\right) }h{\bf \Psi }^{\left(
vac\right) -1}\right] _{\left( o\right) }\left| \lambda _o\right\rangle ,
\label{dr19}
\end{equation}

Using (\ref{dr17}) we obtain 
\begin{equation}
\Theta _{-}^{-1}\left| \lambda _i\right\rangle =\frac{\tau _i(t^{\pm })}{%
\widehat{\tau }_i^{\left( 0\right) }(t^{\pm })}  \label{dr20}
\end{equation}

\section{The $\tau $ function and the N soliton solution for GNLS$_r$}

The GNLS$_r$ model is constructed for example in \ct{fordy}, where it was studied in the framework of the Zakharov-Shabat formalism and the context of hermitian symmetric spaces. In \ct{aratyn} an affine Lie algebraic foundation (loop algebra $g\otimes {\mathbf C} [\lambda,\lambda^{-1}]$ of $g$) for GNLS$_r$ was given.
Here instead we will consider the full Kac-Moody algebra $\widehat{sl}(r+1)$
with homogeneous gradation
\br
s=(1,\stackrel{r-zeros}{\overbrace{0,0,...,0}}),
\er
and a semisimple element $E^{(l)}$.

The connections are given by 
\br
A_1\equiv A&=&E^{\left( 1\right) }+\sum_{i=1}^r\Psi _i^{+}E_{\beta _i}^{\left(
0\right) }+\sum_{i=1}^r\Psi _i^{-}E_{-\beta _i}^{\left( 0\right) }+\nu
_1C,\\
\nonu
A_2\equiv B&=&E^{\left( 2\right) }+\sum_{i=1}^r\Psi _i^{+}E_{\beta _i}^{\left(
1\right) }+\sum_{i=1}^r\Psi _i^{-}E_{-\beta _i}^{\left( 1\right)
}+\sum_{i=1}^r\partial _x\Psi _i^{+}E_{\beta _i}^{\left( 0\right) }-\\
&&\sum_{i=1}^r\partial _x\Psi _i^{-}E_{-\beta _i}^{\left( 0\right)
}-\sum_{i=1}^r\Psi _i^{+}\Psi _i^{-}\beta _i.H^{\left( 0\right)
}-\sum_{i,j=1}^r\Psi _i^{+}\Psi _i^{-}\epsilon \left( \beta _i,\beta
_j\right) E_{\beta _i-\beta _j}^{\left( 0\right) }+\nu _2C,  \label{dr21}
\er
where $\Psi _i^{+}$, $\Psi _i^{-}$, $\nu _1$ and $\nu _2$ are the fields of
the model.

We have
\begin{equation}
\quad E^{\left( l\right) }=2\frac{\mu _r.H^{\left( l\right) }}{\alpha _r^2}%
,\quad \left[ D,E^{\left( l\right) }\right] =lE^{\left( l\right) }
\label{dr22}
\end{equation}
where $\mu _r$ is a fundamental weight and $\alpha _{a}$ are the
simple roots of the $sl(r+1)$ algebra. The $\beta _i$ are the positive roots
defined by
\begin{equation}
\beta _i=\alpha _i+\alpha _{i+1}+...+\alpha _r,\quad \mbox{with}\quad \alpha
_i^2=2,  \label{dr23}
\end{equation}
and $H_i$ being the generators of the cartan subalgebra in the Weyl-Cartan
basis. We need also some relations in the Chevalley basis 
\begin{eqnarray}
E &=&\frac 1{r+1}\left( \sum_{a=1}^ra{\bf H}_a^{\left( o\right) }\right)
,\quad \mbox{with}\quad {\bf H}_a=\alpha _a.H^{\left( o\right) }  \nonumber
\\
\mu _r &=&\frac 1{r+1}\left( \sum_{a=1}^ra\alpha _a\right) ,\quad \beta _a.H=%
{\bf H}_a+...+{\bf H}_r,\quad a=1,...,r  \label{dr24}
\end{eqnarray}
\begin{eqnarray*}
\left[ E^{\left( m\right) },{\bf H}_a^{\left( n\right) }\right] &=&\frac
m{r+1}\sum_{a=1}^ra\eta _{ab}C\delta _{m,-n}, \\
\left[ E^{\left( m\right) },E_{\pm \beta _i}^{\left( n\right) }\right]
&=&\pm E_{\pm \beta _i}^{\left( m+n\right) }, \\
\left[ {\bf H}_a^{\left( m\right) },E_{\pm \beta _i}^{\left( n\right)
}\right] &=&\pm \left( \sum_{b=i}^rK_{ba}\right) E_{\pm \beta _i}^{\left(
m+n\right) },
\end{eqnarray*}
where 
\br
\eta _{ab}=\frac 2{\alpha _2^2}K_{ab}=\eta _{ba}\qquad \mbox{and}\qquad
K_{ab}=2\frac{\alpha _a\cdot \alpha _b}{\alpha _b^2}\cdot 
\er

The potentials are in the subspaces 
\begin{equation}
A_1\,\in\, \widehat{g}_o(s)+\widehat{g}_1(s),\quad A_2\,\in\, \widehat{g}_o(s)+%
\widehat{g}_1(s)+\widehat{g}_2(s)  \label{dr25}
\end{equation}

The zero curvature condition\, $[\partial _t$ $-B$ $,\partial _x$ $-$ $A]=0$\,
gives the following system of equations
\br
\partial _t\Psi _i^{+} &=&\partial _x^2\Psi _i^{+}-2\left( \sum_{j=1}^r\Psi
_j^{+}\Psi _j^{-}\right) \Psi _i^{+},  \nonumber  \\
\label{dr26}
\partial _t\Psi _i^{-} &=&-\partial _x^2\Psi _i^{-}+2\left( \sum_{j=1}^r\Psi
_j^{+}\Psi _j^{-}\right) \Psi _i^{-}, \\
\nonu
\qquad \qquad \qquad \partial _t\nu _1-\partial _x\nu _2 &=&0. 
\er

The system of equations for the $\Psi^{\pm}_{j}$ fields in \rf{dr26}, supplied with a convenient complexification of the time variable and the fields, are related to the well known integrable vector non-linear Schr\"{o}dinger equation (vector {\bf NLS}) \ct{manakov, zen, faddeev}.

Let us now study the vacuum solutions and dressing transformations. Let us
note that $\Psi _i^{\pm }=\nu _1=$ $\nu _2=0$ is a solution of
equations (\ref{dr26}). Therefore from (\ref{dr21}) we have the connections 
\begin{equation}
A_1^{\left( vac\right) }\equiv A^{\left( vac\right) }=E^{\left( 1\right)
},\quad A_2^{\left( vac\right) }\equiv B^{\left( vac\right) }=E^{\left(
2\right) },  \label{dr27}
\end{equation}

These connections can be obtained from\, $A_N^{(vac)}=\partial_{t_N}{\bf \Psi } {\bf \Psi }^{-1}$\, from the group element

\begin{equation}
{\bf \Psi }^{\left( vac\right) }=e^{xE^{\left( 1\right) }+tE^{\left(
2\right) }+X},  \label{dr28}
\end{equation}
where 
\begin{equation}
X=\sum_{n=3}^{+\infty }t_nE^{(n)},\,\,\,t_n\,\,\mbox{ are real parameters.}
\label{dr29}
\end{equation}

The connections in the vacuum orbit are given by 
\begin{eqnarray}
A &=&\Theta _{-}E^{\left( 1\right) }\Theta _{-}^{-1}+\partial _x\Theta
_{-}\Theta _{-}^{-1}\qquad  \label{dr30} \\
&=&M^{-1}NE^{\left( 1\right) }N^{-1}M-M^{-1}\partial _xM+M^{-1}\partial
_xNN^{-1}M,  \nonumber
\end{eqnarray}
\begin{eqnarray}
B &=&\Theta _{-}E^{\left( 2\right) }\Theta _{-}^{-1}+\partial _t\Theta
_{-}\Theta _{-}^{-1}  \label{dr31} \\
&=&M^{-1}NE^{\left( 2\right) }N^{-1}M-M^{-1}\partial _tM+M^{-1}\partial
_tNN^{-1}M,  \nonumber
\end{eqnarray}
with 
\br
\Theta _{-}=\exp \left( \sum_{n>0}\sigma _{-n}\right) ,\quad {\em \ }M={\em %
\exp }\left( \sigma _o\right) ,\quad N=\exp \left( \sum_{n>0}\sigma
_n\right) 
\er

In the present case the gradation operator is\, $Q_s=D$\, with
\begin{equation}
\left[ D,\sigma _n\right] =n\sigma _n.  \label{dr32}
\end{equation}

We can relate the fields $\Psi _i^{\pm }$, $\nu _1$ and $\nu _2$ with some
of the components of $\sigma _n$. We have 
\begin{equation}
A=E^{\left( 1\right) }+\left[ \sigma _{-1},E^{\left( 1\right) }\right] +%
\mbox{ terms of negative grade.\qquad }  \label{dr33}
\end{equation}

\br
=M^{-1}\left( E^{\left( 1\right) }-\partial _xM.M^{-1}+\partial _x\sigma
_1\right) M+\mbox{ terms of grade}>1 
\er

\begin{equation}
B=E^{\left( 2\right) }+\left[ \sigma _{-1},E^{\left( 2\right) }\right]
+\left[ \sigma _{-2},E^{\left( 2\right) }\right] +\frac 12\left[ \sigma
_{-1},\left[ \sigma _{-1},E^{\left( 2\right) }\right] \right] +  \label{dr33.1}
\end{equation}
\[
\mbox{terms of negative grade} 
\]
\[
=M^{-1}\left( E^{\left( 2\right) }-\partial _tM.M^{-1}+\partial _t\sigma
_1+\partial _t\sigma _2+\left[ \sigma _1,\partial _t\sigma _1\right] \right)
M+ 
\]
\[
\mbox{terms of grade }>2 
\]

In (\ref{dr33}) the term of degree $-1$ must vanish and therefore 
\begin{equation}
\partial _x\sigma _{-1}+\left[ \sigma _{-2},E^{\left( 1\right) }\right]
+\frac 12\left[ \sigma _{-1},\left[ \sigma _{-1},E^{\left( 1\right) }\right]
\right] =0,  \label{dr34}
\end{equation}

Denote (consistently with (\ref{dr33})) 
\begin{equation}
\sigma _{-1}=-\sum_{i=1}^r\Psi _i^{+}E_{\beta _i}^{\left( -1\right)
}+\sum_{i=1}^r\Psi _i^{-}E_{-\beta _i}^{\left( -1\right)
}+\sum_{a=1}^r\sigma _{-1}^a{\bf H}_a^{\left( -1\right) },  \label{dr35}
\end{equation}
\[
\sigma _{-2}=\sum_{i=1}^r\sigma _{-2}^{+i}E_{\beta _i}^{\left( -2\right)
}+\sum_{i=1}^r\sigma _{-2}^{-i}E_{-\beta _i}^{\left( -2\right)
}+\sum_{a=1}^r\sigma _{-2}^a{\bf H}_a^{\left( -2\right) }, 
\]
and therefore from (\ref{dr34}) we obtain 
\begin{eqnarray}
\quad \quad \partial _x\sigma _{-1}^a &=&\sum_{i=1}^r\Psi _i^{+}\Psi
_i^{-},\quad a,i=1,...r.  \nonumber \\
\sigma _{-2}^{+i} &=&-\partial _x\Psi _i^{+}+\frac 12\sum_{a=1}^r\sigma
_{-1}^a\Psi _i^{+}\left( \sum_{b=i}^rK_{ba}\right) ,  \label{dr36} \\
\sigma _{-2}^{-i} &=&-\partial _x\Psi _i^{+}+\frac 12\sum_{a=1}^r\sigma
_{-1}^a\Psi _i^{-}\left( \sum_{b=i}^rK_{ba}\right) ,  \nonumber
\end{eqnarray}

Substitution of $\sigma _{-1}$, $\sigma _{-2}$ in (\ref{dr33}) and (\ref{dr33.1}%
) we obtain 
\[
\nu _1=-\frac 1{r+1}\sum_{a,b=1}^ra.\eta _{ab}\sigma _{-1}^b, 
\]
\begin{equation}
\nu _2=-\frac 2{r+1}\sum_{a,b=1}^ra.\eta _{ab}\sigma _{-2}^b.  \label{dr37}
\end{equation}

The $\sigma _{-n}$ 's with the higher gradations are used to cancel out the
undesired componentes.

One of the tau-function vectors is given by 
\begin{equation}
\tau _o\left( x,t\right) =\left[ {\bf \Psi }^{\left( 0\right) }h{\bf \Psi }%
^{\left( 0\right) -1}\right] \left| \lambda _o\right\rangle  \label{dr38}
\end{equation}
\begin{equation}
=\Theta _{-}^{-1}M^{-1}\left| \lambda _o\right\rangle ,  \label{dr39}
\end{equation}
where $h$ is a particular constant element of $\widehat{sl}(r+1)$.

Then we have 
\begin{equation}
\exp \left( -\sum_{n>0}\sigma _{-n}\right) {\em \exp }\left( -\sigma
_o\right) \left| \lambda _o\right\rangle =\left[ {\bf \Psi }^{\left(
0\right) }h{\bf \Psi }^{\left( 0\right) -1}\right] \left| \lambda
_o\right\rangle .  \label{dr40}
\end{equation}

We want to express the fields $\Psi _i^{\pm }$ in terms of some tau
functions, which are some matrix elements in a appropriate representation of 
$\widehat{sl}(r+1)$.

We can write down 
\begin{equation}
\sigma _o=\sum_{a=1}^r\sigma _o^{+a}E_{\alpha _a}^{\left( 0\right)
}+\sum_{a=1}^r\sigma _o^{-a}E_{-\alpha _a}^{\left( 0\right)
}+\sum_{a=1}^r\sigma _o^a{\bf H}_a^{\left( 0\right) }+\eta C  \label{dr41}
\end{equation}
or
\begin{equation}
\sigma _o=\sum_{i=1}^r\sigma _o^{+i}.e_i+\sum_{i=1}^r\sigma
_o^{-i}.f_i+\sum_{a=1}^r\sigma _o^a.h_a+\eta C,  \label{dr42}
\end{equation}
where $e_i$ and $f_i$ ($i=0...r$) are the generators in the Chevalley basis
and \{ $h_i$ , $D$\} generates the Cartan subalgebra.

We have (see Appendix \ref{appa})
\begin{equation}
h_i\left| \lambda _o\right\rangle =0,\quad f_i\left| \lambda _o\right\rangle
=0,\quad e_i\left| \lambda _o\right\rangle =0\quad \mbox{and \quad }C\left|
\lambda _o\right\rangle =\left| \lambda _o\right\rangle ,\quad \quad
i=1,2,...r  \label{dr43}
\end{equation}

Therefore the zero gradation of (\ref{dr40}) is 
\begin{equation}
{\em \exp }\left( -\sigma _o\right) \left| \lambda _o\right\rangle =\left[ 
{\bf \Psi }^{\left( 0\right) }h{\bf \Psi }^{\left( 0\right) -1}\right]
_{\left( o\right) }\left| \lambda _o\right\rangle ,  \label{dr44}
\end{equation}
and the left hand side can be written as 
\begin{equation}
{\em \exp }\left( -\sigma _o\right) \left| \lambda _o\right\rangle =\left|
\lambda _o\right\rangle \widehat{\tau }^{\left( o\right) }\left( x,t\right)
\label{dr45}
\end{equation}
with $\widehat{\tau }^{(0)}$ a real function given by the matrix element 
\begin{equation}
\widehat{\tau }^{\left( o\right) }\left( x,t\right) =\left\langle \lambda
_o\right| \left[ {\bf \Psi }^{\left( 0\right) }h{\bf \Psi }^{\left( 0\right)
-1}\right] _{\left( o\right) }\left| \lambda _o\right\rangle .  \label{dr46}
\end{equation}

Then the term with grade (-1) in (\ref{dr40}) is 
\begin{equation}
-\sigma _{-1}\left| \lambda _o\right\rangle =\frac{\left[ {\bf \Psi }%
^{\left( 0\right) }h{\bf \Psi }^{\left( 0\right) -1}\right] _{\left(
-1\right) }\left| \lambda _o\right\rangle }{\widehat{\tau }^{\left( o\right)
}\left( x,t\right) }  \label{dr47}
\end{equation}
or 
\[
\left( -\sum_{i=1}^r\Psi _i^{+}E_{\beta _i}^{\left( -1\right)
}+\sum_{i=1}^r\Psi _i^{-}E_{-\beta _i}^{\left( -1\right)
}+\sum_{a=1}^r\sigma _{-1}^a{\bf H}_a^{\left( -1\right) }\right) \left|
\lambda _o\right\rangle = 
\]
\begin{equation}
-\frac{\left[ {\bf \Psi }^{\left( 0\right) }h{\bf \Psi }^{\left( 0\right)
-1}\right] _{\left( -1\right) }\left| \lambda _o\right\rangle }{\widehat{%
\tau }^{\left( o\right) }\left( x,t\right) },  \label{dr48}
\end{equation}

Using the commutation rules for the relevant Kac-Moody algebra elements we
have 
\begin{equation}
\Psi _i^{+}=\frac{\tau _i^{+}}{\widehat{\tau }^{\left( o\right) }}\quad 
\mbox{and}\quad \Psi _i^{-}=-\frac{\tau _i^{-}}{\widehat{\tau }^{\left(
o\right) }}\quad ,  \label{dr49}
\end{equation}
where the $tau$ functions are defined by 
\begin{equation}
\tau _i^{+}\equiv \left\langle \lambda _o\right| E_{-\beta _i}^{\left(
1\right) }\left[ {\bf \Psi }^{\left( 0\right) }h{\bf \Psi }^{\left( 0\right)
-1}\right] _{\left( -1\right) }\left| \lambda _o\right\rangle ,  \label{dr50}
\end{equation}
\begin{equation}
\tau _i^{-}\equiv \left\langle \lambda _o\right| E_{\beta _i}^{\left(
1\right) }\left[ {\bf \Psi }^{\left( 0\right) }h{\bf \Psi }^{\left( 0\right)
-1}\right] _{\left( -1\right) }\left| \lambda _o\right\rangle ,  \label{dr51}
\end{equation}
and 
\begin{equation}
\widehat{\tau }^{\left( o\right) }\equiv \left\langle \lambda _o\right|
\left[ {\bf \Psi }^{\left( 0\right) }h{\bf \Psi }^{\left( 0\right)
-1}\right] _{\left( o\right) }\left| \lambda _o\right\rangle .\qquad \quad
\label{dr52}
\end{equation}

In order to obtain the first non trivial solutions of (\ref{dr26}) let us
consider 
\begin{equation}
h=e^{aF_j},\quad F_j=\sum_{n=-\infty }^{+\infty }\nu _j^nE_{-\beta
_j}^{\left( -n\right) },\quad \mbox{where }\nu _j\mbox{ and }a\mbox{ are
real parameters.}  \label{dr53}
\end{equation}
with $F_j$ being an eigenvector under the adjoint action of the generator $E^{(n)}$, that is 
\br
\left[ xE^{\left( 1\right) }+tE^{\left( 2\right) }+X,F_j\right] =-\left[ \nu
_j\left( x+\nu _jt\right) +\overline{\nu }_j\right] F_j,  \label{dr54}
\er
where 
\[
\overline{\nu }_j=\sum_{n=3}^{+\infty }z_n\nu _j^n; 
\]
denoting \,$\varphi _j=\nu _j\left( x+\nu _jt\right) +\overline{\nu }_j$\, one
can write 
\begin{equation}
\left[ {\bf \Psi }^{\left( 0\right) }h{\bf \Psi }^{\left( 0\right)
-1}\right] =\exp \left( e^{-\varphi _j}aF_j\right)  \label{dr55}
\end{equation}
\[
\qquad \qquad \qquad =1+e^{-\varphi _j}aF_j, 
\]
where we have used the fact that $F_j^n=0,$ for $n\geq 2$, that is, $F_j$
are nilpotent (see Appendix \ref{appb} for more details)

The tau function $\widehat{\tau }^{\left( o\right) }$ becomes 
\[
\widehat{\tau }^{\left( o\right) }=\left\langle \lambda _o\right| \left(
1+e^{-\varphi _j}aF_j\right) \left| \lambda _o\right\rangle =1,\quad 
\]
since 
\[
\left\langle \lambda _o\right| E_{-\beta _i}^{\left( o\right) }\left|
\lambda _o\right\rangle =0. 
\]
For the tau function $\tau _i^{+}$ we have 
\begin{eqnarray*}
\tau _i^{+} &=&\left\langle \lambda _o\right| \left[ E_{-\beta _i}^{\left(
1\right) }\exp \left( e^{-\varphi _j}.a\sum_{n=-\infty }^{+\infty }\nu
_j^nE_{-\beta _i}^{\left( -n\right) }\right) \right] _{\left( o\right)
}\left| \lambda _o\right\rangle \\
&=&\left\langle \lambda _o\right| \left[ E_{-\beta _i}^{\left( 1\right)
}e^{-\varphi _j}.a\sum_{n=-\infty }^{+\infty }\nu _j^nE_{-\beta _i}^{\left(
-n\right) }\right] _{\left( o\right) }\left| \lambda _o\right\rangle ,
\end{eqnarray*}
as $\left[ E_{-\beta _i}^{\left( 1\right) },E_{-\beta
_j}^{\left( -n\right) }\right] =0$ \quad  and $\quad E_{-\beta
_i}^{\left( 1\right) }\left| \lambda _o\right\rangle =0,$\,\, it follows

\[
\tau _i^{+}=0. 
\]

Now 
\[
\tau _i^{-}=\left\langle \lambda _o\right| \left[ E_{\beta _i}^{\left(
1\right) }e^{-\varphi _j}.a\sum_{n=-\infty }^{+\infty }\nu _j^nE_{-\beta
_i}^{\left( -n\right) }\right] _{\left( o\right) }\left| \lambda
_o\right\rangle , 
\]
if $i\neq j$ the last expression becomes 
\[
\tau _i^{-}=\left\langle \lambda _o\right| E_{\beta _i-\beta _j}^{\left(
0\right) }e^{-\varphi _j}.a\nu _j\left| \lambda _o\right\rangle =0, 
\]
then 
\[
\tau _i^{-}=a\delta _{ij}e^{-\varphi _j}.\nu _j\left\langle \lambda
_o\right| E_{\beta _i}^{\left( 1\right) }E_{-\beta _i}^{\left( -1\right)
}\left| \lambda _o\right\rangle ; 
\]
the matrix element can be written as 
\[
\left\langle \lambda _o\right| \left\{ \sum_{a=1}^r{\bf H}_a^{\left(
0\right) }+C+E_{-\beta _i}^{\left( -1\right) }E_{\beta _i}^{\left( 1\right)
}\right\} \left| \lambda _o\right\rangle . 
\]
Then it follows 
\[
\tau _i^{-}=\delta _{ij}.a.\nu _{j.}e^{-\varphi _j}, 
\]
and therefore we obtain the solution 
\begin{equation}
\Psi _i^{+}=0,\qquad \Psi _i^{-}=-\delta _{ij}.a.\nu _{j.}e^{-\varphi _j}.
\label{dr6.41}
\end{equation}

We can also make the choice 
\[
h=e^{b.G_j},\mbox{ with }G_j=\sum_{n=-\infty }^{+\infty }\rho _j^nE_{\beta
_i}^{\left( -n\right) },\quad \rho _j\mbox{ being a real
parameter.} 
\]
with $G_j$ a new eigenvector of the adjoint action of $E^{(l)}.$
\[
\left[ xE^{\left( 1\right) }+tE^{\left( 2\right) }+X\,,\,G_j\right] =\left( \rho
_j\left( x+\rho _jt\right) +\overline{\rho }_j\right) G_j, 
\]
with 
\[
\overline{\rho }_j=\sum_{n=3}^{+\infty }z_n\rho _j^n. 
\]

As in the previous case 
\[
\left[ {\bf \Psi }^{\left( 0\right) }h{\bf \Psi }^{\left( 0\right)
-1}\right] =\exp \left( e^{\eta _j}bG_j\right) 
\]
\begin{equation}
\qquad \qquad \qquad =1+e^{\eta _j}bG_j,  \label{dr6.43}
\end{equation}
where $G_j^n=0$ for $n\geq 2$. Denoting\, $\eta _j=\rho _j\left( x+\rho
_jt\right) +\overline{\rho }_j$,\, and performing a similar calculations as in the previous case, one gets
\[
\widehat{\tau }^{\left( o\right) }=1,\qquad \tau _i^{-}=0\quad \mbox{and
\quad }\tau _i^{+}=\delta _{ij}.b.\rho _{j.}e^{\eta _j}, 
\]
and therefore we have a new solution of the form 
\begin{equation}
\Psi _i^{-}=0,\qquad \Psi _i^{+}=\delta _{ij}.b.\rho _{j.}e^{\eta _j}.
\label{dr6.44}
\end{equation}

Consider now the product 
\begin{equation}
h=e^{a_{j_1}F_{j_1}}e^{b_{j_2}G_{j_2}},\quad j_1,j_2=1,2,...r,  \label{6.45}
\end{equation}
where 
\[
F_{j_1}=\sum_{n=-\infty }^{+\infty }\nu _{j_1}^nE_{-\beta _{j_1}}^{\left(
-n\right) },\,\,\,\,G_{j_2}=\sum_{n=-\infty }^{+\infty }\rho
_{j_2}^nE_{\beta _{j_2}}^{\left( -n\right) };\,\,\,\,a_{j_1}, %
b_{j_2}\mbox{ real parameters.} 
\]

Let us note that $F_{j_1}$ and $G_{j_2}$ are associated to positive and
negative roots respectively.

Therefore 
\begin{eqnarray}
\nonu
\left[ {\bf \Psi }^{\left( 0\right) }h{\bf \Psi }^{\left( 0\right)
-1}\right] &=&\left( 1+e^{-\varphi _{j_1}}a_{j_1}F_{j_1}\right) \left(
1+e^{\eta _{j_2}}b_{j_2}G_{j_2}\right) \\
&=&1+e^{-\varphi _{j_1}}a_{j_1}F_{j_1}+e^{\eta
_{j_2}}b_{j_2}G_{j_2}+a_{j_1}b_{j_2}e^{-\varphi _{j_1}}e^{\eta
_{j_2}}F_{j_1}G_{j_2},
\label{dr6.46}
\end{eqnarray}
with $\varphi _{j_1}=\nu_{j_1}\left( x+\nu _{j_1}t\right) +
\overline{\nu }_{j_1}$\quad and \quad $\eta _{j_2}=\rho _{j_2}\left(
x+\rho _{j_2}t\right) +\overline{\rho }_{j_2}$.
The corresponding tau function are 
\begin{equation}
\widehat{\tau }^{\left( o\right) }=1+\delta
_{j_1,j_2}a_{j_1}b_{j_2}C_{j_1,j_2}e^{-\varphi _{j_1}}e^{\eta _{j_2}},\qquad
C_{j_1,j_2}=\frac{\nu _{j_1}.\rho _{j_2}}{\left( \nu _{j_1}-\rho
_{j_2}\right) ^2}\quad ,  \label{dr6.47}
\end{equation}

\begin{equation}
\tau _i^{+}=\left\langle \lambda _o\right| E_{-\beta _i}^{\left( 1\right)
}b_{j_2}e^{\eta _{j_2}}G_{j_2}\left| \lambda _o\right\rangle  \label{dr6.48}
\end{equation}
\[
\quad =\delta _{i,j_2}b_{j_2}\rho _{j_2}e^{\eta _{j_2}},\qquad \qquad 
\]
and 
\begin{equation}
\tau _i^{-}=\delta _{i,j_2}a_{j_1}\nu _{j_1}e^{-\varphi _{j_1}}.\qquad
\qquad \qquad  \label{dr6.49}
\end{equation}

In Appendix \ref{appd} we outline the form of the corresponding matrix elements. We therefore obtain 
\begin{equation}
\Psi _i^{+}=\frac{\delta _{i,j_2}b_{j_2}\rho _{j_2}e^{\eta _{j_2}}}{1+\delta
_{j_1,j_2}a_{j_1}b_{j_2}C_{j_1,j_2}e^{-\varphi _{j_1}}e^{\eta _{j_2}}},\quad
\Psi _i^{-}=-\frac{\delta _{i,j_2}a_{j_1}\nu _{j_1}e^{-\varphi _{j_1}}}{%
1+\delta _{j_1,j_2}a_{j_1}b_{j_2}C_{j_1,j_2}e^{-\varphi _{j_1}}e^{\eta
_{j_2}}}\ \cdot  \label{dr6.50}
\end{equation}

If $j_{1\neq }j_2$ in (\ref{dr6.47}) we recover the solution (\ref{dr6.41}) and (%
\ref{dr6.44}). Therefore in order to have {\sl one-soliton} solutions we must have $%
j_1=j_2=i$ and therefore a solution of the system of equations \rf{dr26} is 
\begin{equation}
\Psi _i^{+}=\frac{b_i\rho _ie^{\eta _i}}{1+a_ib_iC_{i,i}e^{-\varphi
_i}e^{\eta _i}},\quad \Psi _i^{-}=-\frac{a_i\nu _ie^{-\varphi _i}}{%
1+a_ib_iC_{i,i}e^{-\varphi _i}e^{\eta _i}}\quad \cdot  \label{dr6.51}
\end{equation}

In order to study the $N$-soliton solution consider the following group
element 
\begin{equation}
h=e^{a_1F_{i_1}}...e^{a_NF_{i_N}}e^{b_1G_{j_1}}...e^{b_NG_{j_N}},
\label{dr6.52}
\end{equation}
where 
\[
F_{j_l}=\sum_{n=-\infty }^{+\infty }\nu _l^nE_{-\beta _{j_l}}^{\left(
-n\right) },\quad G_{j_l}=\sum_{n=-\infty }^{+\infty }\rho _l^nE_{\beta
_{j_l}}^{\left( -n\right) },\quad l=1,2,...N;\quad \,\,i_l,j_l=1,2,...r; 
\]

\[
\quad \,\,a_l,\quad b_l,\quad \,\,\nu _l\quad \mbox{and \quad }\rho
_l\mbox{ \quad are real parameters.} 
\]
where
\[
\varphi _l=\nu _l\left( x+\nu _lt\right) +\overline{\nu }_l\quad ,\quad \eta
_l=\rho _l\left( x+\rho _lt\right) +\overline{\rho }_l\quad ,l=1,2,...,N. 
\]

Then
\begin{eqnarray}
\left[ {\bf \Psi }^{\left( 0\right) }h{\bf \Psi }^{\left( 0\right)
-1}\right] &=&\left( 1+e^{-\varphi _1}a_1F_{i_1}\right) ...\left(
1+e^{-\varphi _N}a_NF_{i_N}\right) \left( 1+e^{\eta _1}b_1G_{j_1}\right) ...
\nonumber \\
&&\left( 1+e^{\eta _N}b_NG_{j_N}\right) ,  \label{dr6.53}
\end{eqnarray}

Using (\ref{dr50}), (\ref{dr51}) and (\ref{dr52}) we calculate the corresponding tau functions. Denoting 
\begin{equation}
A_{i_l}\equiv a_l\nu _l.e^{-\varphi _l},\quad B_{j_l}\equiv b_{l_l}\rho
_le^{\eta _l},  \label{dr6.54}
\end{equation}
we have 
\[
\widehat{\tau }^{\left( o\right) }=\left\langle \lambda _o\right| \left( 
{\bf \Psi }^{\left( 0\right) }h{\bf \Psi }^{\left( 0\right) -1}\right)
\left| \lambda _o\right\rangle 
\]

\[
\widehat{\tau }^{\left( o\right) }=1+\qquad \qquad {\bf \qquad \quad } 
\]

\br
\nonu
&&\left\langle \lambda _o\right| {\bf \{}\sum_{n=1}^N \sum_{1\leq
l_1<l_2<...<l_N \leq N}\left( A_{i_{l_1}}...A_{i_{l_n}}\right) \left(
F_{i_{l_1}}...F_{i_{l_n}}\right) \sum_{1\leq k_1<k_2<...<k_n\leq N}\left(
B_{j_{k_1}}...B_{j_{k_n}}\right) \\
&&\left( G_{j_{k_1}}...G_{j_{k_n}}\right){\bf \}} \left| \lambda
_o\right\rangle
\er

\br
=1+\sum_{n=1}^N\, \sum_{1\leq l_1<l_2<...<l_n \leq N,\,\, 1\leq k_1<k_2<...<k_n\leq N} C_{i_{l_1}...i_{l_n},
j_{k_1}...j_{k_n}}A_{i_{l_1}}...A_{i_{l_n}}B_{j_{k_1}}...B_{j_{k_n}},
\label{dr6.55}
\er
where the coefficients are given by the matrix elements 
\begin{equation}
C_{i_{l_1}...i_{l_n},j_{k_1}...j_{k_n}}=\left\langle \lambda _o\right|
F_{i_{l_1}}...F_{i_{l_n}}G_{j_{k_1}}...G_{j_{k_n}}\left| \lambda
_o\right\rangle ,  \label{dr6.56}
\end{equation}

Similarly 
\[
\tau _i^{\pm }=\left\langle \lambda _o\right| E_{\mp \beta _i}^{\left(
1\right) }{\bf \Psi }^{\left( 0\right) }h{\bf \Psi }^{\left( 0\right)
-1}\left| \lambda _o\right\rangle 
\]
\br
&=&\left\langle \lambda _o\right| E_{\mp \beta _i}^{\left( 1\right) }\left(
1+e^{-\varphi _1}a_1F_{i_1}\right) ...\left( 1+e^{-\varphi
_N}a_NF_{i_N}\right) \left( 1+e^{\eta _1}b_1G_{j_1}\right) ...  \nonumber \\
&&\left( 1+e^{\eta _N}b_NG_{j_N}\right) \left| \lambda _o\right\rangle ,
\label{dr6.57}
\er

Therefore we obtain 
\br
\tau _i^{+}=\sum_{n=0}^{N-1}\,\,\sum_{1\leq l_1<l_2<...<l_n\leq N,\,\, 1\leq
k_1<k_2<...<k_{n+1}\leq N  } 
C_{ii_{l_1}...i_{l_n},j_{k_1}...j_{k_{n+1}}}^{+}A_{i_{l_1}}...A_{i_{l_n}}B_{j_{k_1}}...B_{j_{k_{n+1}}},
\label{dr6.58}
\er
\br
\tau _i^{-}=\sum_{n=0}^{N-1}\,\,\sum_{ 1\leq l_1<l_2<...<l_{n+1}\leq N,\,\,
1\leq k_1<k_2<...<k_n \leq N }
C_{i_{l_1}...i_{l_{n+1}},ij_{k_1}...j_{k_n}}^{-}A_{i_{l_1}}...A_{i_{l_n}}B_{j_{k_1}}...B_{j_{k_{n+1}}},
\label{dr6.59}
\er
with the matrix elements given by 
\begin{equation}
C_{ii_{l_1}...i_{l_n},j_{k_1}...j_{k_{n+1}}}^{+}=\left\langle \lambda
_o\right| E_{-\beta _i}^{\left( 1\right)
}F_{i_{l_1}}...F_{i_{l_n}}G_{j_{k_1}}...G_{j_{k_{n+1}}}\left| \lambda
_o\right\rangle ,  \label{dr6.60}
\end{equation}

\begin{equation}
C_{i_{l_1}...i_{l_{n+1}},ij_{k_1}...j_{k_n}}^{-}=\left\langle \lambda
_o\right| E_{\beta _i}^{\left( 1\right)
}F_{i_{l_1}}...F_{i_{l_n}}G_{j_{k_1}}...G_{j_{k_n}}\left| \lambda
_o\right\rangle .\qquad  \label{dr6.61}
\end{equation}

The calculation of the corresponding matrix elements is outlined in Appendix \ref{appd}.

\section{The example of GNLS$_1$}

The hierarchy GNLS$_1$ has a Lax operator 
\begin{equation}
L=\partial _x-E^{\left( 1\right) }-\Psi ^{+}E_{+}^{\left( 0\right) }-\Psi
^{-}E_{-}^{\left( 0\right) }-\nu _1C,  \label{7.1}
\end{equation}
where $\Psi ^{\pm }$ and $\nu _1$ are the fields of the model.

The corresponding $\widehat{sl}(2)$ algebra in the Weyl Cartan basis is 
\br
\left[ H^{\left( m\right) },H^{\left( n\right) }\right] &=&\frac n2C\delta
_{m+n,0},  \nonumber \\
\left[ H^{\left( n\right) },E_{\pm }^{\left( m\right) }\right] &=&\pm E_{\pm
}^{\left( m+n\right) },  \label{7.2} \\
\left[ E_{+}^{\left( m\right) },E_{-}^{\left( n\right) }\right]
&=&2H^{\left( m+n\right) }+nC\delta _{m+n,0}.  \nonumber
\er

The equations of the hierarchy are obtained as follows 
\[
\frac{\partial L}{\partial t_N}=\left[ B_N,L\right] ,\qquad N>0 
\]
where 
\[
B_N=\left( UH^{\left( N\right) }U^{-1}\right) _{\geq 0}\in C^\infty \left( 
\IR,\,\widehat{g}_{\geq o}\left( s\right) \right) , 
\]

\[
B_N\subset \bigoplus_{i=0}^N\widehat{g}_i, 
\]
with $U$\ being a group element obtained by exponentiating the negative
degree elements

\[
U=\exp \left( \sum_{n>1}T^{\left( -n\right) }\right) ,\quad \left[
D,T^{\left( n\right) }\right] =nT^{\left( n\right) }. 
\]
The first two $B_N$ are 
\br
B_1&=&H^{\left( 1\right) }+\Psi ^{+}E_{+}^{\left( 0\right) }+\Psi
^{-}E_{-}^{\left( 0\right) }+\nu _1C,  \label{7.3}
\\
B_2 &=&H^{\left( 2\right) }+\Psi ^{+}E_{+}^{\left( 1\right) }+\Psi
^{-}E_{-}^{\left( 1\right) }-2\Psi ^{+}\Psi ^{-}H^{\left( 0\right)
}+\partial _x\Psi ^{+}E_{+}^{\left( 0\right) }- \nonu \\
&&\partial _x\Psi ^{-}E_{-}^{\left( 0\right) }+\nu _2C,  \label{7.5}
\er
and the first equations of the hierarchy are 
\br
\partial _{t_1}L &=&\llbrack B_1, L\rrbrack :\nonu \\
\partial _{t_1}\Psi ^{\pm } &=&\partial _x\Psi ^{\pm }, \label{7.6}\\
\partial _{t_1}\nu _1 &=&\partial _x\nu _1. \nonu
\er
and
\br
\partial _{t_2}L &=&\llbrack B_2,L\rrbrack :  \nonumber \\
\partial _{t_2}\Psi ^{\pm } &=&\pm \partial _x^2\Psi ^{\pm }\mp 2.\left(
\Psi ^{+}\Psi ^{-}\right) \Psi ^{\pm },  \lab{sch} \\
\partial _{t_2}\nu _1 &=&\partial _x\nu _2. \nonu \label{7.8}
\er

The system of equations for the $\Psi^{\pm}$ fields in \rf{sch}, supplied with a convenient complexification of the time variable and the fields, are related to the well known non-linear Schr\"{o}dinger equation ({\bf NLS}) \ct{kac, leznov}. 
The zero curvature condition for $B_1$ and $B_N$ can be written as 
\begin{equation}
\llbrack \partial _{t_N}-B_N,\partial _x-B_1\rrbrack =0,\quad N=1,2,...
\label{7.9}
\end{equation}
where $B_N$ has the general form 
\begin{equation}
B_N=H^{\left( N\right) }+\sum_{n=0}^{N-1}B_N^{\left( n\right) },\quad \mbox{%
with }B_N^{\left( n\right) }\in C^\infty \left( {\bf R}{\em ,}\,\,%
\widehat{g}_n\left( s_{\hom }\right) \right) .  \label{7.10}
\end{equation}

As $\Psi ^{\pm }=\nu _N=0$ is a solution of each system of equations
of the hierarchy, we have 
\begin{equation}
B_1^{\left( vac\right) }=H^{\left( 1\right) },\ \quad B_N^{\left( vac\right)
}=H^{\left( N\right) },  \label{7.11}
\end{equation}
such connections can be obtained with the help of\, $B_N=\partial _{t_N}{\bf %
\Psi \Psi }^{-1}$\, from the group element 
\br
{\bf \Psi }^{\left( vac\right) }=\exp \left( xH^{\left( 1\right)
}+t_NH^{\left( N\right) }+\sum_{ n=2,3,...  n\neq N  } t_nH^{\left(
n\right) }\right) \equiv \exp \left( \sum_{n=1,2,...}t_nH^{\left( n\right)
}\right) .  \label{7.12}
\er

Observe that according to (\ref{7.6}) we have identified $t_1=x.$

The connections in the vacuum orbit are given by 
\begin{equation}
\quad B_1=\Theta H^{\left( 1\right) }\Theta ^{-1}+\partial _x\Theta \Theta
^{-1},\qquad \qquad \qquad \qquad  \label{7.13}
\end{equation}

\[
\quad =M^{-1}\left( {\em N}H^{\left( 1\right) }{\em N}^{-1}-\partial
_xMM^{-1}+\partial _x{\em NN}^{-1}\right) M{\em ,} 
\]
\begin{equation}
\quad \quad B_N=\Theta H^{\left( N\right) }\Theta ^{-1}+\partial
_{t_N}\Theta \Theta ^{-1},\quad \qquad \qquad \qquad \qquad  \label{7.14}
\end{equation}
\[
\quad \quad \quad =M^{-1}\left( {\em N}H^{\left( N\right) }{\em N}%
^{-1}-\partial _{t_N}MM^{-1}+\partial _{t_N}{\em NN}^{-1}\right) M. 
\]

Denote 
\begin{equation}
\Theta =\exp \left( \sum_{n>0}\sigma _{-n}\right) ,\quad M=\exp \left(
\sigma _o\right) ,\quad {\em N}=\exp \left( \sum_{n>0}\sigma _n\right) ,
\label{7.15}
\end{equation}
\[
\llbrack D,\sigma _n\rrbrack =n\sigma _n, 
\]

Therefore we can relate $\Psi ^{\pm }$ to some $\sigma _n$. For instance for 
$N=2$ and denoting $t_2=t$, we have 
\begin{equation}
B_1=H^{\left( 1\right) }+\llbrack \sigma _{-1},H^{\left( 1\right) }\rrbrack +%
\mbox{ terms of negative grade,\qquad \qquad }  \label{7.16}
\end{equation}
\[
=M^{-1}\left( H^{\left( 1\right) }-\partial _xMM^{-1}+\partial _x\sigma
_1\right) M\,\,+\mbox{ terms of grade }>1, 
\]
\[
\quad B_2=H^{\left( 2\right) }+\llbrack \sigma _{-1},H^{\left( 2\right)
}\rrbrack +\llbrack \sigma _{-2},H^{\left( 2\right) }\rrbrack +\frac 12\llbrack
\sigma _{-1},\llbrack \sigma _{-1},H^{\left( 2\right) }\rrbrack \rrbrack + 
\]
\begin{equation}
+\mbox{ terms of negative grade.}  \label{7.17}
\end{equation}
\[
=M^{-1}\left( H^{\left( 2\right) }-\partial _tMM^{-1}+\partial _t\sigma
_1+\partial _t\sigma _2+\llbrack \sigma _1,\partial _t\sigma _1 \rrbrack\right)
M+\mbox{ terms of grade}>2.
\]

Let us observe that the next term (with degree $-1$) in (\ref{7.16})
vanishes, and therefore we have 
\begin{equation}
\partial _x\sigma _{-1}+\llbrack \sigma _{-2},H^{\left( 1\right) }\rrbrack
+\frac 12\llbrack \sigma _{-1},\llbrack \sigma _{-1},H^{\left( 1\right) }\rrbrack
\rrbrack =0.  \label{7.18}
\end{equation}

Denoting 
\br
\sigma _{-1}=-\Psi ^{+}E_{+}^{\left( -1\right) }+\Psi ^{-}E_{-}^{\left(
-1\right) }+\sigma _{-1}^oH^{\left( -1\right) }  \label{7.19}
\er
and 
\br
\sigma _{-2}=-\sigma _{-2}^{+}E_{+}^{\left( -2\right) }+\sigma
_{-2}^{-}E_{-}^{\left( -2\right) }+\sigma _{-2}^oH^{\left( -2\right) },
\label{7.20}
\er
from (\ref{7.18}) we obtain 
\br
\partial _x\sigma _{-1}^o &=&2.\Psi ^{+}\Psi ^{-}  \label{7.21} \\
\sigma _{-2}^{+} &=&-\partial _x\Psi ^{+}+\frac 12\sigma _{-1}^o\Psi ^{+}
\label{7.22} \\
\sigma _{-2}^{-} &=&-\partial _x\Psi ^{-}+\frac 12\sigma _{-1}^o\Psi ^{-}.
\label{7.23}
\er

Substituting these expressions for $\sigma _{-1}$ and $\sigma _{-2}$ in (\ref
{7.16}) and (\ref{7.17} ) we obtain (\ref{7.8} ) with 
\br
\nu _1=-\frac{\sigma _{-1}^o}2,\quad \nu _1=-\sigma _{-2}^o.  \label{7.24}
\er

The $\sigma _{-n}$ 's with higher grades are to cancel the undesired
components. The term of gradation $-2$ in (\ref{7.16}) satisfies \footnote{we use\, $\partial
e^\sigma e^{-\sigma }=\partial \sigma +\frac 1{2!}\left[ \sigma ,\partial
\sigma \right] +\frac 1{3!}\left[ \sigma ,\left[ \sigma ,\partial \sigma
\right] \right] +\cdot \cdot \cdot $} 
\begin{equation}
\partial _x\sigma _{-2}+\llbrack \sigma _{-3},H^{\left( 1\right) }\rrbrack
+\frac 12\llbrack \sigma _{-2},\llbrack \sigma _{-1},H^{\left( 1\right) }\rrbrack
\rrbrack +\frac 12\llbrack \sigma _{-1},\llbrack \sigma _{-2},H^{\left( 1\right)
}\rrbrack \rrbrack =0,  \label{7.25}
\end{equation}
where 
\[
\sigma _{-3}=\sigma _{-3}^{+}E_{+}^{\left( -3\right) }+\sigma
_{-3}^{-}E_{-}^{\left( -3\right) }+\sigma _{-3}^oH^{\left( -3\right) }. 
\]

From (\ref{7.25}) we obtain 
\begin{equation}
\partial _x\sigma _{-2}^o=\Psi ^{-}\partial _x\Psi ^{+}-\Psi ^{+}\partial
_x\Psi ^{-}.  \label{7.26}
\end{equation}

Equating to zero the terms of degree ($-1$) and ($-2$) in (\ref{7.17}) we
can obtain 
\begin{equation}
\partial _t\sigma _{-1}^o=2\left( \Psi ^{-}\partial _x\Psi ^{+}-\Psi
^{+}\partial _x\Psi ^{-}\right) ,  \label{7.27}
\end{equation}
and 
\begin{equation}
\partial _t\sigma _{-2}^o=\frac 23\Psi ^{+}\Psi ^{-}\left( \sigma
_{-1}^o\right) ^2-2\partial _x\Psi ^{+}\partial _x\Psi ^{-}-\frac 23\left(
\Psi ^{+}\Psi ^{-}\right) ^2+\Psi ^{-}\partial _x\Psi ^{+}-\Psi ^{+}\partial
_x\Psi ^{-}.  \label{7.28}
\end{equation}

From (\ref{7.21}) and (\ref{7.27}) we obtain 
\begin{equation}
\partial _t\left( \Psi ^{+}\Psi ^{-}\right) =\partial _xF\llbrack \Psi
^{+},\Psi ^{-}\rrbrack ,  \label{7.29}
\end{equation}
where $F$ is a functional of the fields and ${\cal H}_1$ is the first
Hamiltonian given by

\[
{\cal H}_1=\int_{-\infty }^{+\infty }dx.\Psi ^{+}\Psi ^{-}. 
\]

In the same way (\ref{7.26}) and (\ref{7.28}) gives 
\begin{equation}
\partial _t\left( \Psi ^{-}\partial _x\Psi ^{+}-\Psi ^{+}\partial _x\Psi
^{-}\right) =\partial _xG\llbrack \Psi ^{+},\Psi ^{-}\rrbrack ,  \label{7.30}
\end{equation}
where $G$ is a functional of the fields and ${\cal H}_2$ is the second
Hamiltonian of GNLS$_1$ system given by 
\begin{equation}
{\cal H}_2=\int_{-\infty }^{+\infty }dx.\left( \Psi ^{-}\partial _x\Psi
^{+}-\Psi ^{+}\partial _x\Psi ^{-}\right) .  \label{7.31}
\end{equation}

In this way one can construct the remaining Hamiltonians of higher order corresponding to every $\sigma^{0}_n $ ($n<-2$).

Let us define the tau-function vector
\begin{equation}
\tau \left( x,t_2,t_3,...\right) ={\bf \Psi }^{\left( vac\right) }h{\bf \Psi 
}^{\left( vac\right) -1}\left| \lambda _o\right\rangle ,  \label{7.32}
\end{equation}
\begin{equation}
=\Theta ^{-1}M^{-1}\left| \lambda _o\right\rangle ,  \label{7.33}
\end{equation}
where $h$ is a particular element of the group $\widehat{sl}(2)$ which
generates a dressing transformation.

Therefore we have 
\begin{equation}
\exp \left( -\sum_{n>0}\sigma _{-n}\right) \exp \left( -\sigma _o\right)
\left| \lambda _o\right\rangle ={\bf \Psi }^{\left( vac\right) }h{\bf \Psi }%
^{\left( vac\right) -1}\left| \lambda _o\right\rangle ,  \label{7.34}
\end{equation}
and then one can write
\begin{equation}
\sigma _o=\sigma _o^oH+\sigma _o^{+}E_{+}^{\left( o\right) }+\sigma
_o^{-}E_{-}^{\left( o\right) }+\eta C,  \label{7.35}
\end{equation}
or 
\[
\sigma _o=\sigma _o^oh_1+\sigma _o^{+}e_1+\sigma _o^{-}f_1+\eta C, 
\]
where we have used\, $h_1\left| \lambda _o\right\rangle =0,\quad f_1\left| \lambda
_o\right\rangle =0$\quad and \quad $C\left| \lambda _o\right\rangle =\left|
\lambda _o\right\rangle .$ In this way the zero gradation of expression (\ref
{7.34}) is 
\begin{equation}
\exp \left( -\sigma _o\right) \left| \lambda _o\right\rangle =\left( {\bf %
\Psi }^{\left( vac\right) }h{\bf \Psi }^{\left( vac\right) -1}\right)
_{\left( o\right) }\left| \lambda _o\right\rangle ,  \label{7.36}
\end{equation}
which can be rewritten as
\[
\exp \left( -\sigma _o\right) \left| \lambda _o\right\rangle =\left| \lambda
_o\right\rangle \widehat{\tau }^{\left( o\right) }\left( x,t\right) 
\]
where \,$\widehat{\tau }^{\left( o\right) }\left( x,t\right)$ is a function of 
$x$ and the times $t_n$ given by the following matrix element 
\begin{equation}
\widehat{\tau }^{\left( o\right) }\left( x,t\right) =\left\langle \lambda
_o\right| \left( {\bf \Psi }^{\left( vac\right) }h{\bf \Psi }^{\left(
vac\right) -1}\right) _{\left( o\right) }\left| \lambda _o\right\rangle .
\label{7.37}
\end{equation}

The term with degree ($-1$) in (\ref{7.34}) becomes 
\[
\left( -\sigma _{-1}\right) \left| \lambda _o\right\rangle =\frac{\left( 
{\bf \Psi }^{\left( vac\right) }h{\bf \Psi }^{\left( vac\right) -1}\right)
_{\left( -1\right) }\left| \lambda _o\right\rangle }{\widehat{\tau }^{\left(
o\right) }\left( x,t\right) }, 
\]
or 
\begin{equation}
\left( -\Psi ^{+}E_{+}^{\left( -1\right) }+\Psi ^{-}E_{-}^{\left( -1\right)
}+\sigma _{-1}^oH^{\left( -1\right) }\right) \left| \lambda _o\right\rangle
=-\frac{\left( {\bf \Psi }^{\left( vac\right) }h{\bf \Psi }^{\left(
vac\right) -1}\right) _{\left( -1\right) }\left| \lambda _o\right\rangle }{%
\widehat{\tau }^{\left( o\right) }\left( x,t\right) },  \label{7.38}
\end{equation}

As $H^{\left( 0\right) }\left| \lambda _o\right\rangle =h_1\left| \lambda
_o\right\rangle =0,$ $\quad E_{\pm }^{\left( 1\right) }\left| \lambda
_o\right\rangle =0$ we can write 
\begin{equation}
\Psi ^{+}=\frac{\tau ^{+}}{\widehat{\tau }^{\left( o\right) }}\qquad \mbox{%
and\qquad }\Psi ^{-}=-\frac{\tau ^{-}}{\widehat{\tau }^{\left( o\right) }},
\label{7.39}
\end{equation}
where 
\begin{equation}
\tau ^{+}\equiv \left\langle \lambda _o\right| E_{-}^{\left( 1\right)
}\left( {\bf \Psi }^{\left( vac\right) }h{\bf \Psi }^{\left( vac\right)
-1}\right) _{\left( -1\right) }\left| \lambda _o\right\rangle ,  \label{7.40}
\end{equation}
\begin{equation}
\tau ^{-}\equiv \left\langle \lambda _o\right| E_{+}^{\left( 1\right)
}\left( {\bf \Psi }^{\left( vac\right) }h{\bf \Psi }^{\left( vac\right)
-1}\right) _{\left( -1\right) }\left| \lambda _o\right\rangle .  \label{7.41}
\end{equation}

The relations \rf{7.37}, \rf{7.40} and \rf{7.41} define the tau-functions of the GNLS$_1$ system of equations \rf{sch}.

In order to obtain the first non trivial solution we choose 
\begin{equation}
h=e^F,\qquad \mbox{with\qquad }F=\sum_{n=-\infty }^{+\infty }\nu
_1^nE_{-}^{\left( -n\right) }.  \label{7.42}
\end{equation}

Since 
\begin{equation}
\left[ \sum_{n=1}^{+\infty }t_nH^{\left( n\right) },F\right] =-\left(
\sum_{n=1}^{+\infty }t_n\nu _1^n\right) F,\qquad \mbox{with }\nu _1\mbox{ a
real parameter,}  \label{7.43}
\end{equation}
we may obtain 
\br
\left( {\bf \Psi }^{\left( vac\right) }h{\bf \Psi }^{\left( vac\right)
-1}\right) &=&\exp \left( e^{-\varphi _1}F\right) \\
&=&1+e^{-\varphi _1}F,
\er
with 
\[
\varphi _1=\sum_{n=1}^{+\infty }t_n\nu _1^n. 
\]
Where we have used the property $F^n=0,$ for $n\geq 2$. The tau functions
become 
\[
\tau ^{\left( o\right) }=\left\langle \lambda _o\right| \left( 1+e^{-\varphi
_1}E_{-}^{\left( o\right) }\right) \left| \lambda _o\right\rangle =1, 
\]
\[
\quad \quad \tau ^{+}=\left\langle \lambda _o\right| E_{-}^{\left( 1\right)
}e^{-\varphi _1}\nu _1E_{-}^{\left( -1\right) }\left| \lambda
_o\right\rangle =0, 
\]

\br
\tau ^{-} &=&\left\langle \lambda _o\right| E_{+}^{\left( 1\right)
}e^{-\varphi _1}\nu _1E_{-}^{\left( -1\right) }\left| \lambda _o\right\rangle
\\
&=&\nu _1e^{-\varphi _1}\left\langle \lambda _o\right| \left( 2H^{\left(
0\right) }+C\right) \left| \lambda _o\right\rangle \\
&=&\nu _1e^{-\varphi _1}.
\er

Using (\ref{7.39}) we obtain the following solution of the Eqs. \rf{sch} 
\begin{equation}
\Psi ^{+}=0\qquad \mbox{e \quad }\Psi ^{-}=-\nu _1e^{-\varphi _1}.
\label{7.44}
\end{equation}
Now let us choose 
\begin{equation}
h=e^G,\qquad G=\sum_{n=-\infty }^{+\infty }\rho _1^nE_{+}^{\left( -n\right)
},\,\,\rho _1\,\mbox{is a real parameter.}  \label{7.45}
\end{equation}
Since 
\[
\llbrack \sum_{n=1}^{+\infty }t_nH^{\left( n\right) },G\rrbrack =\left(
\sum_{n=1}^{+\infty }t_n\rho _1^n\right) G, 
\]
we obtain 
\br
\left( {\bf \Psi }^{\left( vac\right) }h{\bf \Psi }^{\left( vac\right)
-1}\right) &=&\exp \left( e^{\eta _1}G\right) \\
&=&1+e^{\eta _1}G,\qquad \mbox{with\qquad }\eta _1=\sum_{n=1}^{+\infty
}t_n\rho _1^n,
\er
where we used $G^n=0,$ $n\geq 2.$ Therefore the tau functions become
\[
\widehat{\tau }^{\left( o\right) }=\left\langle \lambda _o\right| \left(
1+e^{\eta _1}E_{+}^{\left( o\right) }\right) _{\left( o\right) }\left|
\lambda _o\right\rangle =1, 
\]

\[
\tau ^{-}=\left\langle \lambda _o\right| \left( E_{+}^{\left( 1\right)
}e^{\eta _1}\rho _1E_{+}^{\left( -1\right) }\right) _{\left( o\right)
}\left| \lambda _o\right\rangle =0, 
\]
\[
\tau ^{+}=\left\langle \lambda _o\right| E_{-}^{\left( 1\right) }e^{\eta
_1}\rho _1E_{+}^{\left( -1\right) }\left| \lambda _o\right\rangle =\rho
_1e^{\eta _1}, 
\]
and the corresponding solutions 
\begin{equation}
\Psi ^{-}=0\qquad \mbox{and\qquad }\Psi ^{+}=\rho _1e^{\eta _1}.
\label{7.46}
\end{equation}

In order to obtain {\sl one-soliton} solutions, let us choose

\br
h &=&e^{aF}e^{bG},\qquad \mbox{where }F\mbox{ and }G\mbox{ are given in (\ref
{7.42}) and (\ref{7.45})}  \label{7.47} \\
&&\qquad \qquad \qquad \mbox{with }a\mbox{ and }b\mbox{ real parameters.} 
\nonumber
\er
Then 
\br
{\bf \Psi }^{\left( vac\right) }h{\bf \Psi }^{\left( vac\right) -1} &=&\exp
\left( e^{-\varphi }aF\right) \exp \left( e^\eta bG\right)  \label{7.48} \\
&=&1+e^{-\varphi }aF+e^\eta bG+e^{-\varphi }e^\eta aFbG,  \label{7.49}
\er
with $\varphi =\sum_{n=1}^{+\infty }t_n\nu ^n$ \quad and $\quad \eta
=\sum_{n=1}^{+\infty }t_n\rho ^n.$ $\widehat{\tau }^{\left( o\right) }.$ Let
us compute the relevant tau functions. The expression for $\widehat{\tau }%
^{\left( o\right) }$ becomes

\begin{equation}
\widehat{\tau }^{\left( o\right) }=1+a\,b\,c\,e^{-\varphi }e^\eta ,  \label{7.50}
\end{equation}
where c is an matrix element of the following form 
\br
c &=&\left\langle \lambda _o\right| \left( FG\right) _{\left( o\right)
}\left| \lambda _o\right\rangle  \nonumber  \label{7.2.20} \\
&=&\left\langle \lambda _o\right| \left( \sum_{n,m>0}^{+\infty }\nu
_1^{-n}\rho _1^mE_{-}^{\left( n\right) }E_{+}^{\left( -m\right) }\right)
_{\left( o\right) }\left| \lambda _o\right\rangle  \nonumber \\
&=&\left\langle \lambda _o\right| \sum_{n,m>0}^{+\infty }\nu _1^{-n}\rho
_1^m\left( -2H^{\left( n-m\right) }+m\delta _{n-m,o}C\right) _{\left(
o\right) }\left| \lambda _o\right\rangle  \nonumber \\
&=&\sum_{n=0}^{+\infty }n\left( \frac{\rho _1}{\nu _1}\right) ^n,  \nonumber
\\
&=&\frac{\nu _1\rho _1}{\left( \rho _1-\nu _1\right) ^2}\cdot
\label{7.51}
\er
where, for pedagogical reasons, we have used step by step, the properties of the integrable highest weight representation of the algebra $\hat{sl}(2)$, see Appendix \ref{appa} (Eqs. \rf{a29}-\rf{a35}). The computation of higher order matrix elements is performed very quickly using the vertex operator formalism, see Appendix \ref{appe}.

The remaining tau functions are given by 
\br
\tau ^{+} &=&\left\langle \lambda _o\right| E_{-}^{\left( 1\right) }\left(
be^\eta \rho _1E_{+}^{\left( -1\right) }\right) \left| \lambda
_o\right\rangle  \nonumber \\
&=&be^\eta \rho _1\left\langle \lambda _o\right| E_{-}^{\left( 1\right)
}\left( -2H^{\left( 0\right) }+C\right) \left| \lambda _o\right\rangle 
\nonumber \\
&=&b\rho _1e^\eta  \label{7.52}
\er

and 
\br
\quad \quad \tau ^{-} &=&\left\langle \lambda _o\right| E_{+}^{\left(
1\right) }\left( a.e^{-\varphi }\nu _1E_{-}^{\left( -1\right) }\right)
\left| \lambda _o\right\rangle \qquad \qquad  \nonumber \\
&=&a.\nu _1e^{-\varphi }.  \label{7.53}
\er
Thus, a {\sl one-soliton} solution is 
\begin{equation}
\Psi ^{+}=\frac{b\rho _1e^\eta }{1+a\,b\,c\,e^{-\varphi }e^\eta },  \label{7.54}
\end{equation}
and 
\begin{equation}
\Psi ^{-}=-\frac{a\,\nu _1e^{-\varphi }}{1+a\,b\,c\,e^{-\varphi }e^\eta }.
\label{7.55}
\end{equation}

Let us write down the explicit form of this {\sl one-soliton} solution. As a
particular case we set the following relations

\br
\rho _1=-\nu _1,\quad b=-a=-2\quad \mbox{and \quad }t_{2n+1}=0\quad (n\geq
0), 
\er
then the relations (\ref{7.54}) and (\ref{7.55}) become 
\begin{equation}
\Psi ^{+}=\nu _1\exp \left( \nu _1^2t+\overline{\nu }_1\right) \mbox{sech} \left(
\nu _1x\right)  \label{7.57}
\end{equation}
and 
\begin{equation}
\quad \Psi ^{-}=-\nu _1\exp \left( -\nu _1^2t-\overline{\nu }_1\right) \mbox{sech} \left( \nu _1x\right) ,\quad  \label{7.58}
\end{equation}
where $\overline{\nu }_1=\sum_{n=2}^{+\infty }\nu _1^{2n}t_{2n}$ is a phase
parameter as far as only the equation \rf{sch} of the whole hierarchy is considered. The
solutions of types \rf{7.57} and (\ref{7.58}), are known as an `envelope soliton' or `bright soliton' solutions in the context of the non-linear Schr\"{o}dinger equation \ct{drazin}.

Next let us choose

\br
h &=&e^{a_1F_1}\ e^{a_2F_2}\,\,e^{b_1G_1}\,\,e^{b_2G_2},\quad
a_i,b_i,\nu _i,\rho _i\mbox{ real parameters;}  \nonumber  \label{7.4.1} \\
\mbox{ with }\,F_i &=&\sum_{n=-\infty }^{+\infty }\nu _i^nE_{-}^{\left(
-n\right) }\quad \mbox{and }G_i=\sum_{n=-\infty }^{+\infty }\rho
_i^nE_{+}^{\left( -n\right) },\,\,i=1,2.  \label{7.59}
\er
Then 
\br
&&{\bf \Psi }^{\left( vac\right) }h{\bf \Psi }^{\left( vac\right)
-1}=1+e^{-\varphi _1}a_1F_1+e^{-\varphi _2}a_2F_2+e^{\eta _1}b_1G_1+e^{\eta
_2}b_2G_2+\nonu\\
&&e^{\eta _1}e^{\eta _2}b_1G_1b_2G_2+e^{-\varphi _1}e^{-\varphi
_2}a_1F_1a_2F_2+e^{-\varphi _1}e^{\eta _1}a_1F_1b_1G_1+e^{-\varphi
_1}e^{\eta _2}a_1F_1b_2G_2+\nonu\\
&&e^{-\varphi _2}e^{\eta _2}a_2F_2b_2G_2+e^{\eta _1}e^{-\varphi
_2}b_1G_1a_2F_2+e^{\eta _1}e^{-\varphi _2}e^{\eta _2}b_1G_1a_2F_2b_2G_2+ \nonu\\
&&e^{-\varphi _1}e^{-\varphi _2}e^{\eta _2}a_1F_1a_2F_2b_2G_2+e^{-\varphi
_1}e^{\eta _1}e^{-\varphi _2}a_1F_1b_1G_1a_2F_2+\nonu\\
&&e^{-\varphi _1}e^{\eta _1}e^{\eta _2}a_1F_1b_1G_1b_2G_2+e^{-\varphi
_1}e^{\eta _1}e^{-\varphi _2}e^{\eta _2}a_1F_1b_1G_1a_2F_2b_2G_2,
\label{7.60}
\er
with 
\[
\varphi _i=\sum_{n=1}^{+\infty }\nu _i^nt_n,\quad \eta
_i=\sum_{n=1}^{+\infty }\rho _i^nt_n,\quad i=1,2. 
\]
The corresponding tau functions are 
\br
&&
\widehat{\tau }^{(o)}=1+ e^{-\varphi _1}e^{\eta
_1}a_1b_1\left\langle \lambda _o\right| F_1G_1\left| \lambda _o\right\rangle
+e^{-\varphi _1}e^{\eta _2}a_1b_2\left\langle \lambda _o\right| F_1G_2\left|
\lambda _o\right\rangle +\nonu\\
&& e^{-\varphi _2}e^{\eta
_2}a_2b_2\left\langle \lambda _o\right| F_2G_2\left| \lambda _o\right\rangle
+e^{\eta _1}e^{-\varphi _2}b_1a_2\left\langle \lambda _o\right| G_1F_2\left|
\lambda _o\right\rangle + \nonu\\
&&e^{-\varphi _1}e^{\eta _1}e^{-\varphi _2}e^{\eta
_2}a_1b_1a_2b_2\left\langle \lambda_o\right| F_1G_1F_2G_2\left| \lambda
_o\right\rangle ,  \label{7.61}
\er

\br
\tau ^{+} &=&e^{\eta _1}b_1\left\langle \lambda _o\right| E_{-}^{\left(
1\right) }G_1\left| \lambda _o\right\rangle +e^{\eta _2}b_2\left\langle
\lambda _o\right| E_{-}^{\left( 1\right) }G_2\left| \lambda _o\right\rangle +
\nonumber \\
&&e^{-\varphi _1}e^{\eta _1}e^{\eta _2}a_1b_1b_2\left\langle \lambda
_o\right| E_{-}^{\left( 1\right) }F_1G_1G_2\left| \lambda _o\right\rangle + 
\nonumber \\
&&e^{\eta _1}e^{-\varphi _2}e^{\eta _2}b_1a_2b_2\left\langle \lambda
_o\right| E_{-}^{\left( 1\right) }G_1F_2G_2\left| \lambda _o\right\rangle
\label{7.62}
\er
and 
\br
\tau ^{-} &=&e^{-\varphi _1}a_1\left\langle \lambda _o\right| E_{+}^{\left(
1\right) }F_1\left| \lambda _o\right\rangle +e^{-\varphi _2}a_2\left\langle
\lambda _o\right| E_{+}^{\left( 1\right) }F_2\left| \lambda _o\right\rangle +
\nonumber \\
&&e^{-\varphi _1}e^{-\varphi _2}e^{\eta _2}a_1a_2b_2\left\langle \lambda
_o\right| E_{+}^{\left( 1\right) }F_1F_2G_2\left| \lambda _o\right\rangle + 
\nonumber \\
&&e^{-\varphi _1}e^{\eta _1}e^{-\varphi _2}a_1b_1a_2\left\langle \lambda
_o\right| E_{+}^{\left( 1\right) }F_1G_1F_2\left| \lambda _o\right\rangle .
\label{7.63}
\er
Notice that only some terms of the expansion ${\bf \Psi }^{\left(
vac\right) }h{\bf \Psi }^{(vac)-1}$ contribute to the tau functions. For
example the terms $\left\langle \lambda _o\right| F_i\left| \lambda
_o\right\rangle ,$ $\left\langle \lambda _o\right| G_i\left| \lambda
_o\right\rangle ,$ $\left\langle \lambda _o\right| F_iG_jF_k\left| \lambda
_o\right\rangle $ and $\left\langle \lambda _o\right| G_iF_jG_k\left|
\lambda _o\right\rangle $ do not contribute to the computation of $\widehat{%
\tau }^{\left( o\right) }$, since these matrix elements vanish.

In particular let us consider $\rho _i=-\nu _i,$ $b_i=-a_i=-2$ and $t_{2n+1}=0,$ then

\begin{equation}
\tau ^{+}=a_1\nu _1e^{\widehat{\varphi }_1}+a_2\nu _2e^{\widehat{\varphi }%
_2}+a_1a_2^2\Delta _1e^{\widehat{\varphi }_1}e^{-\varphi _2}e^{\widehat{%
\varphi }_2}+a_1^2a_2\Delta _2e^{-\varphi _1}e^{\widehat{\varphi }_1}e^{%
\widehat{\varphi }_2},\qquad \qquad  \label{7.65}
\end{equation}
\begin{equation}
\tau ^{-}=a_1\nu _1e^{-\varphi _1}+a_2\nu _2e^{-\varphi _2}+a_1a_2^2\Delta
_1e^{-\varphi _1}e^{-\varphi _2}e^{\widehat{\varphi }_2}+a_1^2a_2\Delta
_2e^{-\varphi _1}e^{\widehat{\varphi }_1}e^{-\varphi _2},\quad  \label{7.66}
\end{equation}
where 
\br
\widehat{\varphi }_i &=&-\nu _i\left( x-\nu _it\right) +\overline{\nu }_i, \\
\varphi _i &=&\nu _i\left( x+\nu _it\right) +\overline{\nu }_i,
\er
\[
\Delta _i=\frac{\nu _i}4\left( \frac{\nu _1-\nu _2}{\nu _1+\nu _2}\right)
^2,\quad i=1,2; 
\]
now let us choose $a_1$ and $a_2$ such that 
\[
\nu _i=a_j^2\Delta _i,\qquad \left( i\neq j\right) , 
\]
then 
\[
a_1=a_2\equiv a=2\left( \frac{\nu _1-\nu _2}{\nu _1+\nu _2}\right) . 
\]
Let us remark that the parameters $\overline{\nu }_i$ are some phase
parameters if only the equations \rf{sch} of the hierarchy are to be
considered. With this choice of parameters the $\widehat{\tau }^{\left(
o\right) }$ function turns out to be 
\br
\widehat{\tau }^{\left( o\right) } &=&e^{-\left( \nu _1+\nu _2\right)
x}\{\frac 4{a^2}(e^{\left( \nu _1-\nu _2\right) x}+e^{-\left( \nu _1-\nu
_2\right) x})+e^{\left( \nu _1+\nu _2\right) x}+e^{-\left( \nu _1+\nu
_2\right) x}+  \nonumber \\
&&4\frac{\nu _1\nu _2}{\left( \nu _1-\nu _2\right) ^2}(e^{\left( \nu
_1^2-\nu _2^2\right) t+\left( \overline{\nu }_1-\overline{\nu }_2\right)
}+e^{-\left( \nu _1^2-\nu _2^2\right) t-\left( \overline{\nu }_1-\overline{%
\nu }_2\right) }\}.  \label{7.67}
\er
Therefore the fields $\Psi ^{+}$and $\Psi ^{-}$ become 
\[
\Psi ^{\pm }=\pm ae^{\pm \left[ \left( \nu _1^2+\nu _2^2\right) t+\left( 
\overline{\nu }_1+\overline{\nu }_2\right) \right] }\cdot 
\]
\begin{equation}
\frac{e^{\mp \left( \nu _1^2t+\overline{\nu }_1\right) }\nu _2\cosh \left(
\nu _1x\right) +e^{\mp \left( \nu _2^2t+\overline{\nu }_2\right) }\nu
_1\cosh \left( \nu _2x\right) }{\frac 4{a^2}\cosh \left[ \left( \nu _1-\nu
_2\right) x\right] +\cosh \left[ \left( \nu _1+\nu _2\right) x\right] +4%
\frac{\nu _1\nu _2}{\left( \nu _1-\nu _2\right) ^2}\cosh \left[ \left( \nu
_1^2-\nu _2^2\right) t+\left( \overline{\nu }_1-\overline{\nu }_2\right)
\right] },  \label{7.68}
\end{equation}
which are the {\sl two-soliton} solutions of \rf{sch} or the {\sl two-soliton} solutions (after relevant complexification) of the corresponding non-linear Schr\"{o}dinger equation \ct{twosoliton}.

Regarding the solutions of the equations of higher order of the hierarchy (\ref{7.9}), we may argue that the same solutions, 1-soliton \rf{7.54}-\rf{7.55} and 2-soliton constructed  with the tau functions \rf{7.61}, \rf{7.62}
and \rf{7.63} satisfying \rf{sch} should satisfy the higher order
equations of the hierarchy GNLS$_1,$ each equation with its corresponding time scale $t_n.$ This behaviour is also observed in the study of the KdV system
using the perturvative reduction and multiple time scaling approach \ct{kraenkel}.  

\subsection{N-soliton solutions}

The generalization for a N-soliton solution can be made choosing

\begin{equation}
h=e^{a_1F_1}\,\,...e^{a_NF_N}\,\,e^{b_1G_1}\,\,%
...\,\,e^{b_NG_N},  \label{7.69}
\end{equation}
where 
\br
F_i &=&\sum_{n=-\infty }^{+\infty }\nu _i^nE_{-}^{\left( -n\right) },\quad
G_i=\sum_{n=-\infty }^{+\infty }\rho _i^nE_{+}^{\left( -n\right) },\,\,i=1,2,...,N \\
&&\mbox{with }a_i,b_i,\nu _i\mbox{ and }\rho _i\mbox{ are real parameters .}
\er
The following expression plays an important role in the construction af the
tau functions 
\begin{equation}
{\bf \Psi }^{\left( vac\right) }h{\bf \Psi }^{\left( vac\right) -1}=\left(
1+a_1e^{-\varphi _1}F_1\right) \left( 1+b_1e^{\eta _1}G_1\right) ...\left(
1+a_Ne^{-\varphi _N}F_N\right) \left( 1+b_Ne^{\eta _N}G_N\right) ,
\label{7.70}
\end{equation}
where 
\[
\varphi _i=\sum_{n=1}^{+\infty }\nu _i^nt_n,\quad \eta
_i=\sum_{n=1}^{+\infty }\rho _i^nt_n,\quad i=1,2,...,N; 
\]
the parameters $\nu _i$ and $\rho _i$ will characterize each soliton.

It will be convenient to write the various $F_i$ and $G_i$ in terms of the
vertex operators (see Appendix \ref{appc})
\begin{equation}
F_i\longrightarrow \nu _i\Gamma _{-}\left( \nu _i\right) ,\qquad
G_i\longrightarrow \rho _i\Gamma _{+}\left( \rho _i\right) ,  \label{7.71}
\end{equation}
then 
\[
{\bf \Psi }^{\left( vac\right) }h{\bf \Psi }^{\left( vac\right) -1}=\left(
1+a_1\nu _1e^{-\varphi _1}\Gamma _{-}\left( \nu _1\right) \right) \left(
1+b_1\rho _1e^{\eta _1}\Gamma _{+}\left( \rho _1\right) \right) ... 
\]
\[
\left( 1+a_N\nu _Ne^{-\varphi _N}\Gamma _{-}\left( \nu _N\right) \right)
\left( 1+b_N\rho _Ne^{\eta _N}\Gamma _{+}\left( \rho _N\right) \right) . 
\]
Denoting 
\[
A_n\equiv a_n\nu _ne^{-\varphi _n},\quad B_n\equiv b_n\rho _ne^{\eta _n}, 
\]
we may compute the relevant tau functions 
\[
\widehat{\tau }^{\left( o\right) }=1+\left\langle \lambda _o\right|
\sum_{n=1}^N 
\]
\[
\sum_{1\leq i_1<i_2<...<i_n\leq N}A_{i_1}...A_{i_n}\Gamma _{-}\left( \nu
_{i_1}\right) ...\Gamma _{-}\left( \nu _{i_n}\right) \sum_{1\leq
j_1<j_2<...<j_n\leq N}B_{j_1}...B_{j_n}\Gamma _{+}\left( \rho _{j_1}\right)
...\Gamma _{-}\left( \rho _{j_n}\right) \left| \lambda _o\right\rangle . 
\]

We are using some properties of the product of vertex operators acting on $%
\left| \lambda _o\right\rangle $ and its dual $\left\langle \lambda
_o\right| $ (see Appendix \ref{appb} and \ref{appc}). Then 
\br
\widehat{\tau }^{\left( o\right) } &=&1+\sum_{n=1}^N\,\,\sum_{ 1\leq
i_1<i_2<...<i_n\leq N,\,\, 1\leq j_1<j_2<...<j_n\leq N}
A_{i_1}...A_{i_n}B_{j_1}...B_{j_n}\cdot \\
&&\llbrack \prod_{1\leq l<m\leq n}\left( \nu _{i_l}-\nu _{i_m}\right) ^2\left(
\rho _{j_l}-\rho _{j_m}\right) ^2\epsilon \left( \alpha _{i_l},\alpha
_{i_m}\right) \epsilon \left( \alpha _{j_l},\alpha _{j_m}\right) \rrbrack
\cdot \\
&&\llbrack \prod_{1\leq l\leq m<n}\left( \nu _{i_l}-\rho _{j_m}\right)
^2\epsilon \left( \alpha _{i_l},\alpha _{j_m}\right) \rrbrack ^{-1}.
\er

Denoting 
\[
\epsilon \left( \alpha _{i_l},\alpha _{i_m}\right) =\epsilon \left( -\alpha
,-\alpha \right) \equiv \epsilon \left( -,-\right) , 
\]
\[
\epsilon \left( \alpha _{j_l},\alpha _{j_m}\right) =\epsilon \left( \alpha
,\alpha \right) \equiv \epsilon \left( +,+\right) , 
\]
\[
\epsilon \left( \alpha _{i_l},\alpha _{j_m}\right) =\epsilon \left( -\alpha
,\alpha \right) \equiv \epsilon \left( -,+\right) , 
\]
and using the properties of the cocycles in the case of $\widehat{sl}(2)$ \ct{goddard} 
\br
\epsilon \left( +,+\right) &=&\epsilon \left( -,-\right) =-1, \\
\epsilon \left( +,-\right) &=&\epsilon \left( -,+\right) =1,
\er
we can write 
\br
\widehat{\tau }^{\left( o\right) } &=&1+\sum_{n=1}^N\,\,\sum_{1\leq
i_1<i_2<...<i_n\leq N,\,\, 1\leq j_1<j_2<...<j_n\leq N}
A_{i_1}...A_{i_n}B_{j_1}...B_{j_n}\cdot \\
&&\llbrack \prod_{1\leq l<m\leq n}\left( \nu _{i_l}-\nu _{i_m}\right) ^2\left(
\rho _{j_l}-\rho _{j_m}\right) ^2\rrbrack \llbrack \prod_{1\leq l\leq m\leq
n}\left( \nu _{i_l}-\rho _{j_m}\right) ^2 \rrbrack ^{-1}
\er
and 
\br
\tau ^{\pm } &=&\frac 1{2\pi i}\oint dz.z\left\langle \lambda _o\right|
\Gamma _{\mp }\left( z\right) \left( 1+a_1\nu _1e^{-\varphi _1}\Gamma
_{-}\left( \nu _1\right) \right) \left( 1+b_1\rho _1e^{\eta _1}\Gamma
_{+}\left( \rho _1\right) \right) \cdot \cdot \cdot \\
&&\left( 1+a_N\nu _Ne^{-\varphi _N}\Gamma _{-}\left( \nu _N\right) \right)
\left( 1+b_N\rho _Ne^{\eta _N}\Gamma _{+}\left( \rho _N\right) \right)
\left| \lambda _o\right\rangle .
\er
According to Eq \ref{33}, in order to have non vanishing terms, we must have equal number of operators $\Gamma _{+}$ and $\Gamma _{-}$
inside the states $\left\langle \lambda _o\right| $ and $\left| \lambda
_o\right\rangle $, then 
\br
\tau ^{\pm } &=&\frac 1{2\pi i}\oint
dz.z\sum_{n=1}^NA_1^{m_1}...A_N^{m_N}B_1^{n_1}...B_N^{n_N}\left\langle
\lambda _o\right| \Gamma _{\mp }\left( z\right) \Gamma _{-}^{m_1}\left( \nu
_1\right) ...\Gamma _{-}^{m_N}\left( \nu _N\right) \cdot \\
&&\Gamma _{+}^{n_1}\left( \rho _1\right) ...\Gamma _{+}^{n_N}\left( \rho
_N\right) \left| \lambda _o\right\rangle ,
\er
where the exponents satisfy the following relations 
\br
\sum_{i=1}^Nm_i\pm 1 &=&\sum_{i=1}^Nn_i=n \\
m_i,n_i &=&0,1.
\er
Reordering the operators we may write 
\br
\tau ^{+} &=&\frac 1{2\pi i}\oint d\nu .\nu \sum_{n=0}^{N-1}\,\,\sum_{1\leq
i_1<i_2<...<i_n\leq N,\,\, 1\leq j_1<j_2<...<j_{n+1}\leq N}
A_{i_1}...A_{i_n}B_{j_1}...B_{j_{n+1}}\\
&&\left\langle \lambda _o\right| \Gamma _{-}\left( \nu \right) \Gamma
_{-}\left( \nu _{i_1}\right) ...\Gamma _{-}\left( \nu _{i_n}\right) \Gamma
_{+}\left( \rho _{j_1}\right) ...\Gamma _{+}\left( \rho _{j_{n+1}}\right)
\left| \lambda _o\right\rangle ,
\er
\br
&=&\frac 1{2\pi i}\sum_{n=0}^{N-1}\,\,\sum_{ 1\leq i_1<i_2<...<i_n\leq N\,\, 
1\leq j_1<j_2<...<j_{n+1}\leq N} \oint d\nu .\nu \left(
\prod_{0<l\leq n}\epsilon \left( -\alpha ,\alpha _{i_l}\right) \left( \nu
-\nu _{i_l}\right) ^2\right) \\
&&\left( \prod_{0<m\leq n+1}\epsilon \left( -\alpha ,\alpha _{j_m}\right)
\left( \nu -\rho _{j_m}\right) ^2\right) ^{-1}
A_{i_1}...A_{i_n}B_{j_1}...B_{j_{n+1}} \\
&&\left( \prod_{0\leq l<m\leq n}\epsilon \left( \alpha _{i_l},\alpha
_{i_m}\right) \left( \nu _{i_l}-\nu _{i_m}\right) ^2\right) \cdot \left(
\prod_{0\leq l<m\leq n+1}\epsilon \left( \alpha _{j_l},\alpha _{j_m}\right)
\left( \rho _{j_l}-\rho _{j_m}\right) ^2\right) \\
&&\left( \prod_{ 1\leq l\leq m\leq n+1,\, l\neq n+1 } \epsilon
\left( \alpha _{i_l},\alpha _{j_m}\right) \left( \nu _{i_l}-\rho
_{j_m}\right) ^2\right) ^{-1}\cdot
\er

In Appendix \ref{appf} we show that the contour integration in the variable 
$\nu $ is equal to $2\pi i$ for any value of $n$. Then, 
\begin{eqnarray*}
\tau ^{+} &=&\sum_{n=0}^{N-1}\sum_{ 1\leq i_1<i_2<...<i_n\leq N,\, 1\leq
j_1<j_2<...<j_{n+1}\leq N} A_{i_1}...A_{i_n}B_{j_1}...B_{j_{n+1}}%
\left( \prod_{1\leq l<m\leq n}\epsilon \left( \alpha _{i_l},\alpha
_{i_m}\right) \left( \nu _{i_l}-\nu _{i_m}\right) ^2\right) \cdot \\
&&\left( \prod_{1\leq l<m\leq n+1}\epsilon \left( \alpha _{j_l},\alpha
_{j_m}\right) \left( \rho _{j_l}-\rho _{j_m}\right) ^2\right) .\left(
\prod_{0<l\leq n}\epsilon \left( -\alpha ,\alpha _{i_l}\right) \right) \cdot
\\
&&\left( \prod_{1\leq l\leq m\leq n+1}\epsilon \left( \alpha _{i_l},\alpha
_{j_m}\right) \left( \nu _{i_l}-\rho _{j_m}\right) ^2\right) ^{-1}\left(
\prod_{0<m\leq n+1}\epsilon \left( -\alpha ,\alpha _{j_m}\right) \right)
^{-1}\cdot
\end{eqnarray*}

Likewise we have 
\begin{eqnarray*}
\tau ^{-} &=&\frac 1{2\pi i}\oint d\rho .\rho \sum_{n=0}^{N-1}\,\,\sum_{ 1\leq
i_1<i_2<...<i_n\leq N,\,\, 1\leq j_1<j_2<...<j_{n+1}\leq N}
B_{i_1}...B_{i_n}A_{j_1}...A_{j_{n+1}}. \\
&&\left\langle \lambda _o\right| \Gamma _{+}\left( \rho \right) \Gamma
_{+}\left( \rho _{i_1}\right) ...\Gamma _{+}\left( \rho _{i_n}\right) \Gamma
_{-}\left( \nu _{j_1}\right) ...\Gamma _{-}\left( \nu _{j_{n+1}}\right)
\left| \lambda _o\right\rangle
\end{eqnarray*}
\quad 
\begin{eqnarray*}
&=&\frac 1{2\pi i}\sum_{n=0}^{N-1}\,\,\sum_{1\leq i_1<i_2<...<i_n\leq N,\,\,
1\leq j_1<j_2<...<j_{n+1}\leq N} \oint d\rho .\rho \left(
\prod_{0<l\leq n}\epsilon \left( \alpha ,\alpha _{i_l}\right) \left( \rho
-\rho _{i_l}\right) ^2\right) \\
&&\left( \prod_{0<m\leq n+1}\epsilon \left( \alpha ,\alpha _{j_m}\right)
\left( \rho -\nu _{j_m}\right) ^2\right) ^{-1}
B_{i_1}...B_{i_n}A_{j_1}...A_{j_{n+1}} \\
&&\left( \prod_{1\leq l<m\leq n}\epsilon \left( \alpha _{i_l},\alpha
_{i_m}\right) \left( \rho _{i_l}-\rho _{i_m}\right) ^2\right) \cdot \left(
\prod_{1\leq l<m\leq n+1}\epsilon \left( \alpha _{j_l},\alpha _{j_m}\right)
\left( \nu _{j_l}-\nu _{j_m}\right) ^2\right) \\
&&\left( \prod_{ 1\leq l\leq m\leq n+1 ,\, l\neq n+1} \epsilon
\left( \alpha _{i_l},\alpha _{j_m}\right) \left( \rho _{i_l}-\nu
_{j_m}\right) ^2\right) ^{-1},
\end{eqnarray*}
the contour integration in $\rho $ is also equal to $2\pi i$ ( see Appendix \ref{appf}) for any value of $n.$ Therefore 
\begin{eqnarray*}
\tau ^{-} &=&\sum_{n=0}^{N-1}\sum_{ 1\leq i_1<i_2<...<i_n\leq N,\, 1\leq
j_1<j_2<...<j_{n+1}\leq N} B_{i_1}...B_{i_n}A_{j_1}...A_{j_{n+1}}%
\left( \prod_{1\leq l<m\leq n}\epsilon \left( \alpha _{i_l},\alpha
_{i_m}\right) \left( \rho _{i_l}-\rho _{i_m}\right) ^2\right) \cdot \\
&&\left( \prod_{1\leq l<m\leq n+1}\epsilon \left( \alpha _{j_l},\alpha
_{j_m}\right) \left( \nu _{j_l}-\nu _{j_m}\right) ^2\right) \left(
\prod_{0<l\leq n}\epsilon \left( \alpha ,\alpha _{i_l}\right) \right) \cdot
\\
&&\left( \prod_{1\leq l\leq m\leq n+1}\epsilon \left( \alpha _{i_l},\alpha
_{j_m}\right) \left( \rho _{i_l}-\nu _{j_m}\right) ^2\right) ^{-1}\left(
\prod_{0<m\leq n+1}\epsilon \left( \alpha ,\alpha _{j_m}\right) \right)
^{-1}\cdot
\end{eqnarray*}

Similar expressions were found by Kac and Wakimoto \ct{wakimoto} in the context of their generalized Hirota equations approach.

\section{Acknowledgements} I am grateful to Profs. L.A. Ferreira and J.F. Gomes for enlightening discussions, and Prof. A.H. Zimerman for suggesting me to study this problem and valuable discussions. I thank Prof. B.M. Pimentel for his encouragement. This work was supported by FAPESP under grant 96/00212-0.

\appendix

\section{Appendix: The ``untwisted'' Kac-Moody algebras and their integrable highest-weight representations}
\label{appa}

We present the necessary Kac-Moody algebra notations and conventions used to construct integrable models, as well as, the theory of the so called ``highest weight integrable representations'' which are useful in the construction of their soliton solutions. A complete treatment can be found in \ct{kac, wan}.

An ``untwisted'' Kac-Moody algebra $\widehat{g}$ affine to a finite Lie algebra $g$ can be realized as an extension of the ``loop algebra'' of  $g$:

\begin{equation}
\widehat{g}=\left( g\otimes {\mathbf C} \llbrack z,z^{-1}\rrbrack \right) \oplus 
{\mathbf C} C\oplus {\mathbf C} D,  \lab{a1}
\end{equation}
where ${\mathbf C}  \llbrack z,z^{-1}\rrbrack$ is the algebra of Laurent Polynomials in $z$ and  $\mathbf{C}$$x$ ($x=C,D$) is a 1 dimensional subspace. Writing an element of the loop algebra as $a_{n\equiv }$ $\left( a\otimes z^m\right) ,$ where $a\in g$ and  $%
n\in \mathbf{Z,}$ then the algebra can be written as
\br
\llbrack a_n,b_m\rrbrack \,=\,\llbrack a,b\rrbrack _{n+m}+\delta _{m+n,0}\left(
a,b\right) mC,\nonu\\
\llbrack D,a_n\rrbrack\, =\, na_n \lab{algebra} \\
\llbrack D,D\rrbrack=\llbrack C,D\rrbrack =\llbrack C,C\rrbrack =\llbrack C,a_n\rrbrack
=0, 
\er
where $\left( a,b\right) $ is the Killing form of  $g$ and  $\left[
a,b\right] $ is the Lie bracket in $g$. $C$ is the central element of $\widehat{g}$, and $D$ is a derivative operator which induces a natural integer gradation of  $\widehat{g}$

\[
\widehat{g}=\bigoplus_{i\in \mathbf{Z}}\widehat{g}_i, 
\]
$\ $where $\left[ D,\widehat{g}_i\right] =i\widehat{g}_i.$ The operator $D$
defines the so called  ``homogeneous gradation''.

Here let us point out that there is a natural Heisenberg subalgebra of $\widehat{g}$. Introducing a triangular decomposition of the finite algebra $g=n_{-}\oplus h\oplus n_{+}$, one can define an homogeneous  Heisenberg algebra as an algebra composed of the elements \{$h\otimes z^n,C$\} such that

\[
\llbrack a_n,b_m\rrbrack =\delta _{m+n,0}\left( a,b\right) mC. 
\]

In the Cartan-Weyl basis the commutation relations are given as

\br
\llbrack H_i^m,H_j^n\rrbrack =mC\delta _{ij}\delta _{m,-n},  \lab{a2}
\er
\br
\llbrack H_i^m,E_\alpha ^n\rrbrack =\alpha _iE_\alpha ^{m+n}, \lab{a3}
\er
\br
\llbrack E_\alpha ^m,E_\beta ^n\rrbrack =\left\{ 
\begin{tabular}{l}
$\varepsilon (\alpha ,\beta )E_{\alpha +\beta }^{m+n}\quad ,\mbox{if }\,\alpha
+\beta\,\,\mbox{is a root}$ \\ 
$\frac 2{\alpha ^2}\alpha \cdot H^{m+n}+Cm\delta _{m+n,0}, \quad
\quad \mbox{if }\alpha +\beta =0$ \\ 
$0\quad \quad \quad \quad \quad \quad ,\mbox{in other case}$%
\end{tabular}
\right.  \lab{a4}
\er
\br
\llbrack C,E_\alpha ^m\rrbrack =\llbrack C,H_\alpha ^m\rrbrack =0  \lab{a5}
\er
\br
\llbrack D,E_\alpha ^n\rrbrack =nE_\alpha ^n,\qquad \llbrack D,H_i^n\rrbrack
=nH_i^n.  \lab{a6}
\er

In this case the Cartan subalgebra is formed by the generators \{$%
H_i^{(0)},C,D$\} and the step operators are:\\ 
- $E_\alpha ^n$ associated to the roots $a=\left( \alpha ,0,n\right) ,$ where $\alpha $ belongs to the set of roots of $g$ and $n$ are integers,\\
- $H_i^n$ associated to the roots  $n\delta =\left( 0,0,n\right) $ with $n\neq 0.$

The positive roots are  $\left( \alpha ,0,n\right) $ for $n>0$ or  
 $n=0$ and  $\alpha >0$, and among them the simple roots being  $a_i=\left(
\alpha _i,0,0\right),\,\quad i=1,...r$,\, and  $\alpha _o=\left( -\psi ,0,1\right)
 $  with $\psi$ the maximal root of $g$.

The representation theory of  the Kac-Moody algebra in terms of vertex operators usually make use of the Cartan-Weyl basis \ct{goddard}, see Appendix \ref{appb}. Instead, to construct the so called ``integrable highest weight representations'' \ct{kac, wan} of the ``untwisted'' affine Kac-Moody algebra we will need the Chevalley basis commutation relations (the notations and presentation here follow closely the Appendix of \ct{ferreira1})

\br
&&\llbrack \emph{H}_a^m,\emph{H}_b^n\rrbrack =mC\eta _{ab}\delta _{m+n,0}  \lab{a7}\\
&&\llbrack \emph{H}_a^m,E_\alpha ^n\rrbrack =\sum_{b=1}^rm_b^\alpha
K_{ba}E_\alpha ^{m+n} 
\lab{a8}\\
&&\llbrack E_\alpha ^m,E_{-\alpha }^n\rrbrack
=\sum_{a=1}^rl_a^\alpha \emph{H}_a^{m+n}+\frac 2{\alpha ^2}mC\delta
_{m+n,0}  \lab{a9}
\\
&&\left[ E_\alpha ^m,E_\beta ^n\right] =\left( q+1\right)
\varepsilon (\alpha ,\beta )E_{\alpha +\beta }^{m+n};\qquad \mbox{se }\alpha
+\beta \mbox{ \'{e} uma raiz\qquad }  \lab{a10}
\\
&&\left[ C,E_\alpha ^m\right] =\left[ C,\emph{H}_\alpha ^m\right]
=0 \lab{a11}
\\
&&\left[ D,E_\alpha ^n\right] =nE_\alpha ^n,\qquad \left[ D,\emph{H}%
_a^n\right] =n\emph{H}_a^n.  \lab{a12}
\er
where $K_{ab}=2\alpha _a\cdot \alpha _b/\alpha _b^2$\, is the Cartan matrix of the finite simple Lie algebra  $g$ associated to $\widehat{g}$ and generated by \{$\emph{H}_a^0,E_\alpha ^0$\}. $\eta _{ab}=\frac 2{\alpha
_a^2}K_{ab}=\eta _{ba},$ $q$ is the highest positive integer such that $\beta
-q\alpha $ is a root, $\varepsilon (\alpha ,\beta )$ are $(\pm)$ signs determined by the Jacobi identities, $l_a^\alpha $ and  $m_a^\alpha $ are the integers in the expansion $\alpha /\alpha
^2=\sum_{a=1}^rl_a^\alpha \alpha _a/\alpha _a^2$ and $\alpha
=\sum_{a=1}^rm_a^\alpha \alpha _a$ respectively, where $\alpha _1,...,\alpha _r$ are the simple roots of $g\,(r\equiv \mbox{rank of}\,  g).$ $\widehat{g}$ 
has a symmetric non-degenerate bilinear form which can be normalized as
\[
Tr\left( \emph{H}_a^m\emph{H}_b^n\right) =\eta _{ab}\delta _{m+n,0}
\]
\[
\qquad \qquad Tr\left( E_\alpha ^mE_\beta ^n\right) =\frac 2{\alpha
^2}\delta _{\alpha +\beta ,0}\delta _{m+n,0}\quad 
\]
\[
Tr\left( CD\right) =1.\quad 
\]
The integer gradations of $\widehat{g}$ 
\[
\widehat{g}=\bigoplus_{n\in \mathbf{Z}}\widehat{g}_n
\]
have been presented in \ct{kac}. The gradation operator $Q_s$ satisfying 
\begin{equation}
\left[ Q_{\mathbf{s}},\widehat{g}_n\right] =n\widehat{g}_n;\qquad n\in 
\mathbf{Z,}  \lab{a13}
\end{equation}
is defined by
\begin{equation}
Q_{\mathbf{s}}=H_{\mathbf{s}}+N_{\mathbf{s}}D+\sigma C,\quad H_{\mathbf{s}%
}=\sum_{a=1}^rs_a\lambda _a^v\cdot H^0,\quad H^0=\left(
H_1^0,...H_r^0\right)   \lab{a14}
\end{equation}
where $\left( s_o,s_1,...,s_r\right) $ is a $n-$tuple of non-negative co-prime integers, and  $\lambda _a^v\equiv 2\lambda _a/\alpha _a^2$
with $\lambda _a$ and  $\alpha _a$ being the fundamental weights and the simple roots of $g$ respectively. Moreover, 
\begin{equation}
N_s=\sum_{i=0}^rs_im_i^\psi ,\quad \psi =\sum_{a=1}^rm_a^\psi \alpha
_a,\quad m_0^\psi \equiv 1,  \lab{a15}
\end{equation}
with $\psi $ being the maximal root of  $g$. The value of  $\sigma $ is arbitrary. Therefore 
\[
\quad \left[ Q_{\mathbf{s}},\emph{H}_a^n\right] =nN_{\mathbf{s}}\emph{H}%
_a^n\qquad \qquad \qquad \qquad 
\]
\[
\left[ Q_{\mathbf{s}},E_\alpha ^n\right] =\left( \sum_{a=1}^rm_a^\alpha
s_a+nN_{\mathbf{s}}\right) E_\alpha ^n.
\]
The positive and negative ``step operators'' of $\widehat{g}$ associated to the simple roots are: 
\begin{equation}
e_a\equiv E_{\alpha _a}^0,\quad e_o\equiv E_{-\psi }^1,\quad f_a\equiv
E_{-\alpha _a}^o\quad \mbox{e \quad }f_o\equiv E_\psi ^{-1},  \lab{a16}
\end{equation}
and its Cartan subalgebra generated by
\begin{equation}
h_a\equiv \emph{H}_a^0,\quad h_0\equiv -\sum_{a=1}^rl_a^\psi \emph{H}%
_a^0+\frac 2{\psi ^2}C\mbox{ \quad e \quad }D,  \lab{a17}
\end{equation}
with $l_a^\psi$ given above; then they satisfy 
\begin{equation}
\left[ Q_{\mathbf{s}},h_i\right] =\left[ Q_{\mathbf{s}},D\right] =0;\quad
\left[ Q_{\mathbf{s}},e_i\right] =s_ie_i;\qquad \left[ Q_{\mathbf{s}%
},f_i\right] =-s_if_i;\quad i=0,1,...,r.  \lab{a18}
\end{equation}

An important class of representations of the Kac-Moody algebra are the so called ``integrable highest-weight representations'' \ct{kac}. They are defined in terms of a highest weight $\left| \lambda _{\mathbf{s}}\right\rangle $ labelled by the gradation $\mathbf{s}$ of $\widehat{g}.$  That state is annihilated by the positive grade generators 
\begin{equation}
\widehat{g}_{+}\left| \lambda _{\mathbf{s}}\right\rangle =0,  \lab{a19}
\end{equation}
and it is an eigenstate of the generators of the subalgebra $\widehat{g}_o$ 
\begin{equation}
h_i\left| \lambda _{\mathbf{s}}\right\rangle =s_i\left| \lambda _{\mathbf{s}%
}\right\rangle \qquad \qquad \qquad \qquad \qquad \quad \quad   \lab{a20}
\end{equation}
\begin{equation}
f_i\left| \lambda _{\mathbf{s}}\right\rangle =0;\quad \quad \mbox{for any }i\mbox{ with }s_i=0  \lab{a21}
\end{equation}
\begin{equation}
Q_s\left| \lambda _{\mathbf{s}}\right\rangle =\eta _s\left| \lambda _{%
\mathbf{s}}\right\rangle \qquad \qquad \qquad \qquad \qquad \qquad 
\lab{a22}
\end{equation}
\begin{equation}
\quad C\left| \lambda _{\mathbf{s}}\right\rangle =\frac{\psi ^2}2\left(
\sum_{i=0}^rl_i^\psi s_i\right) \left| \lambda _{\mathbf{s}}\right\rangle
,\qquad \qquad \qquad \quad   \lab{a23}
\end{equation}
where $l_i^\psi $ is given by 
\begin{equation}
\frac \psi {\psi ^2}=\sum_{a=1}^rl_a^\psi \frac{\alpha _a}{\alpha _a^2}%
;\quad l_0^\psi =1.  \lab{a24}
\end{equation}
The eigenvalue of the central element $C$ in the representation $\left|
\lambda _{\mathbf{s}}\right\rangle ,$ is known as the level of the representation 
\begin{equation}
c=\frac{\psi ^2}2\left( \sum_{i=0}^rl_i^\psi s_i\right) , \lab{a25}
\end{equation}
in particular, the ``highest-weight integrable representations'' with $c=1$ are known as ``basic representations''.

The states of highest weight $\left| \lambda _{\mathbf{s}}\right\rangle $ can be realized as 

\begin{equation}
\left| \lambda _{\mathbf{s}}\right\rangle \equiv \bigotimes_{i=0}^r\left| 
\widehat{\lambda }_i\right\rangle ^{\otimes s_i},  \lab{a26}
\end{equation}
where $\left| \widehat{\lambda }_i\right\rangle $ are the highest states of the fundamental representations of $\widehat{g},$ and  $
\widehat{\lambda }_i$ are the relevant fundamental weights of $\widehat{g},$ . They are given by \ct{goddard}
\begin{equation}
\widehat{\lambda }_0=\left( 0,\frac{\psi ^2}2,0\right) \qquad  \lab{a27}
\end{equation}
\begin{equation}
\widehat{\lambda }_a=\left( \lambda _a,\frac{l_a^\psi \psi ^2}2,0\right) ,
\lab{a28}
\end{equation}
where $\lambda _a$, $a=1,2...,r$ are the fundamental weights of the finite Lie algebra $g$ associated to $\widehat{g}$, $l_a^\psi $ is defined in 
\ref{a24}, and the corresponding components are the eigenvalues of $\emph{H}_a^0,$ $C$ and $%
D$ respectively, viz. 
\begin{equation}
\emph{H}_a^0\left| \widehat{\lambda }_0\right\rangle =0;\quad C\left| 
\widehat{\lambda }_0\right\rangle =\frac{\psi ^2}2\left| \widehat{\lambda }%
_0\right\rangle  \lab{a29}
\end{equation}
\begin{equation}
\emph{H}_b^0\left| \widehat{\lambda }_a\right\rangle =\delta _{a,b}\left| 
\widehat{\lambda }_a\right\rangle ;\quad C\left| \widehat{\lambda }%
_a\right\rangle =\frac{\psi ^2}2l_a^\psi \left| \widehat{\lambda }%
_0\right\rangle  \lab{a30}
\end{equation}
and
\begin{equation}
D\left| \widehat{\lambda }_i\right\rangle =0.  \lab{a31}
\end{equation}
Let us notice that in each of the $r+1$ fundamental representations of $\widehat{g}$, the (unique) highest weight state satisfies
\begin{equation}
h_j\left| \widehat{\lambda }_i\right\rangle =\delta _{ij}\left| \widehat{%
\lambda }_i\right\rangle  \lab{a32}
\end{equation}
\begin{equation}
e_j\left| \widehat{\lambda }_i\right\rangle =0,\,\forall \,j
\lab{a33}
\end{equation}
\begin{equation}
\qquad \qquad f_j\left| \widehat{\lambda }_i\right\rangle =0,\mbox{ \quad
for }j\neq i  \lab{a34}
\end{equation}
\begin{equation}
f_j^2\left| \widehat{\lambda }_i\right\rangle =0.\qquad  \lab{a35}
\end{equation}

Then the generators  $e_i$ and $f_{i}$ are nilpotent when acting on $\left| \lambda _{\mathbf{s}}\right\rangle ,$ and these are indeed, integrable representations.

\section{\bf The construction of the homogeneous vertex operators}
\label{appb}
The construction of vertex operator representations of Kac-Moody algebras can be found in \ct{kac, vertex}. The construction of the homogeneous vertex operators is based on the homogeneous Heisenberg subalgebra\,${\mathbf h}$. Here we follow the construction developed in \ct{kac}.

Consider

\begin{equation}
g={\bf h}\bigoplus \left( \bigoplus_{\alpha \in \Delta }{\bf C}E_\alpha
\right) ,  \label{1}
\end{equation}
$g$ being a simple finite Lie algebra of type $A_l,D_l$ or $E_l$, whose
commutation relations are given by 
\br
\llbrack h,h^{,}\rrbrack =0,\qquad \mbox{ \qquad \qquad if }h,\,h^{,}\in 
{\bf h\qquad } 
\er
\br
\quad \quad \llbrack h,E_\alpha \rrbrack =\left( h\mid \alpha \right) E_\alpha ,%
\mbox{ \quad \quad if }h\in {\bf h,}\quad \alpha \in \Delta \quad 
\er
\[
\llbrack E_\alpha ,E_{-\alpha }\rrbrack =-\alpha ,\qquad \qquad \qquad \mbox{if 
}\alpha \in \Delta \qquad \qquad 
\]
\[
\quad \quad \llbrack E_\alpha ,E_\beta \rrbrack =0,\qquad \mbox{if }\alpha
,\beta \in \Delta ,\quad \alpha +\beta \notin \Delta \cup \left\{ 0\right\} 
\]
\[
\llbrack E_\alpha ,E_\beta \rrbrack =\varepsilon \left( \alpha ,\beta \right)
E_{\alpha +\beta },\qquad \mbox{if }\alpha ,\beta ,\alpha +\beta \in \Delta
, 
\]
where $\Delta =\left\{ \alpha \in Q/(\alpha \mid \alpha )=2\right\} $ and $Q$
is the root lattice; $\left( \quad \mid \quad \right) $ is the invariant
symmetric form of $g,$ normalized as follows$:$ 
\[
(h\mid E_\alpha )=0,\mbox{ if }h\in {\bf h,}\quad \alpha \in \Delta ;\qquad
(E_\alpha ,E_\beta )=-\delta _{\alpha ,-\beta },\mbox{ if }\alpha ,\beta \in
\Delta . 
\]
Let

\begin{equation}
\widehat{g}={\bf C}\left[ t,t^{-1}\right] \otimes _{{\bf C}}g+{\bf C}C+{\bf C%
}D,  \label{2}
\end{equation}
be an affine algebra of type $A_n^{\left( 1\right) },D_n^{\left( 1\right) }$
or $E_n^{\left( 1\right) },$ respectively.

Consider the complex commutative associative algebra 
\begin{equation}
V=S\left( \bigoplus_{j<0}(t^j\otimes {\bf h})\right) \otimes _{{\bf C}}{\bf C%
}\left[ Q\right] ,  \label{3}
\end{equation}
where $S$ stands for the symmetric algebra and ${\bf C}\left[ Q\right] $
stands for the group algebra of the root lattice $Q\subset {\bf h}$ of $g$.
Let $\alpha $ $\rightarrow e^\alpha $ denote the inclusion $Q\subset {\bf C}%
\left[ Q\right] $(a base of the vector space ${\bf C}\left[ Q\right] $ is
given by the elements $e^\alpha ,$ $\alpha \in Q,$ and, it is defined the
twisted product of the group algebra elements, $e^\alpha e^\beta
=\varepsilon \left( \alpha ,\beta \right) $ $e^{\alpha +\beta }$ \ct{vandeleur}); $%
u^{\left( n\right) }$ will stand for $t^n\otimes u$ ($n\in {\bf Z},$ $u\in g$%
). For $n>0$, $u\in {\bf h,}$ denote by $u\left( -n\right) $ the operator on 
$V$ of multiplication by $u^{\left( -n\right) }$ $.$ For $n\geq 0$, $u\in 
{\bf h,}$ denote by $u\left( n\right) $ the derivation of the algebra $V$
defined by the formula

\begin{equation}
u(n)\left( v^{\left( -m\right) }\otimes e^\alpha \right) =n\delta
_{n,-m}\left( u\mid v\right) \otimes e^\alpha +\delta _{n,0}\left( \alpha
\mid u\right) v^{\left( -m\right) }\otimes e^\alpha  \label{4}
\end{equation}

Choosing dual bases $u_i$ and $u^i$ of ${\bf h}$, define the operator $D_o$
on $V$ by the formula 
\begin{equation}
D_o=\sum_{i=1}^l\left( \frac 12u_i(0)u^i(0)+\sum_{n\geq
1}u_i(-n)u^i(n)\right) .  \label{5}
\end{equation}
Furthermore, for $\alpha \in Q,$ define the sign operator $c_\alpha $: 
\begin{equation}
c_\alpha \left( f\otimes e^\beta \right) =\varepsilon (\alpha ,\beta
)f\otimes e^\beta .  \label{6}
\end{equation}
Finaly, for $\alpha \in \Delta \subset Q$ introduce the {\bf vertex operator}
\begin{equation}
\Gamma _\alpha (z)=\exp \left( \sum_{j\geq 1}\frac{z^j}j\alpha (-j)\right)
\exp \left( -\sum_{j\geq 1}\frac{z^{-j}}j\alpha (j)\right) e^\alpha
z^{\alpha (0)}c_\alpha ,  \label{7}
\end{equation}
here $z$ is viewed as an indeterminate. Expanding in powers of $z$: 
\begin{equation}
\Gamma _\alpha (z)=\sum_{j\in {\bf Z}}\Gamma _\alpha ^{\left( j\right)
}z^{-j-1},  \label{8}
\end{equation}
we obtain a sequence of operators $\Gamma _\alpha ^{\left( j\right) }$ on $V 
$. Now we can state the result.

{\bf Theorem}: The map $\sigma :\widehat{g}\rightarrow End\left( V\right) $
defined by 
\begin{eqnarray}
C &\longrightarrow &1,  \nonumber \\
u^{\left( n\right) } &\longrightarrow &u\left( n\right) ,\qquad \mbox{for }%
u\in {\bf h,\qquad n}\in {\bf Z,}  \nonumber \\
E_\alpha ^{\left( n\right) } &\longrightarrow &\Gamma _\alpha ^{\left(
n\right) },\qquad \mbox{for }\alpha \in \Delta {\bf ,\qquad n}\in {\bf Z,} 
\nonumber \\
D &\longrightarrow &-D_o,  \label{9}
\end{eqnarray}
defines the basic representation of the affine algebra $\widehat{g}$ on $V$. 
$End\left( V\right) $ denotes the space of the linear maps of an vector
space $V$ on itself.

The proof of this theorem is presented in \ct{kac}.

\section{\bf Homogeneous vertex operator calculus}
\label{appc}
Defining

\begin{equation}
\Gamma _\alpha ^{\pm }(z)=\exp \sum_{j\geq 1}\frac{\alpha \left( \pm
j\right) }{\mp j}z^{\mp j},\quad \Gamma _\alpha ^0(z)=e^\alpha z^{\alpha
\left( 0\right) }c_\alpha ,  \label{19}
\end{equation}
the following relations can be obtained: 
\begin{equation}
\Gamma _\alpha ^{-}(z_1)\Gamma _\beta ^{+}(z_2)=\Gamma _\beta
^{+}(z_2)\Gamma _\alpha ^{-}(z_1)\left( 1-\frac{z_2}{z_1}\right) ^{\left(
\alpha \mid \beta \right) }  \label{20}
\end{equation}
and 
\begin{equation}
\Gamma _\alpha ^0(z_1)\Gamma _\beta ^0(z_2)=e^{\alpha +\beta }z_1^{\alpha
\left( 0\right) }z_2^{\beta \left( 0\right) }z_1^{\left( \alpha \mid \beta
\right) }c_\alpha c_\beta \varepsilon \left( \alpha ,\beta \right) ,\quad
\label{21}
\end{equation}
where $\left( 1-\frac{z_2}{z_1}\right) ^m,$ $m\in {\bf Z,}$ with $\mid \frac{%
z_2}{z_1}\mid $ $\leq 1.$

From (\ref{20}) and (\ref{21}) it follows 
\begin{eqnarray}
\Gamma _\alpha (z_1)\Gamma _\beta (z_2) &=&\left( 1-\frac{z_2}{z_1}\right)
^{\left( \alpha \mid \beta \right) }z_1^{\left( \alpha \mid \beta \right)
}\varepsilon (\alpha ,\beta )\exp \left[ \sum_{j\geq 1}\frac 1j\left(
z_1^j\alpha (-j)+z_2^j\beta (-j)\right) \right]  \nonumber \\
&&\exp \left[ -\sum_{j\geq 1}\frac 1j\left( z_1^{-j}\alpha (j)+z_2^{-j}\beta
(j)\right) \right] e^{\alpha +\beta }z_1^{\alpha \left( 0\right) }z_2^{\beta
\left( 0\right) }c_\alpha c_\beta .  \label{22}
\end{eqnarray}
If $z\equiv z_1=z_2\quad $and$\quad \alpha =\beta ,$ from (\ref{22}) we can
obtain
\[
\Gamma _\alpha (z)\Gamma _\alpha (z)=0, 
\]
or 
\begin{equation}
\left[ \Gamma _\alpha (z)\right] ^n=0,\qquad \mbox{for\quad }n\geq 2.
\label{23}
\end{equation}
{\bf A useful formula} 
\[
\Gamma _{\alpha _N}(z_N)\Gamma _{\alpha _{N-1}}(z_{N-1})...\Gamma _{\alpha
_1}(z_1)\left( 1\otimes 1\right) =\prod_{1\leq i<j\leq N}\varepsilon (\alpha
_i,\alpha _j)\left( z_i-z_j\right) ^{\left( \alpha _i\mid \alpha _j\right)
}\cdot 
\]
\begin{equation}
\left( \prod_{i=1}^N\exp \sum_{j\in {\bf N}}\frac{z_i^j}j\alpha
_i(-j)\right) \otimes \exp \left( \sum_{i=1}^N\alpha _i\right)  \label{24}
\end{equation}

{\bf Proof}.: We will prove by induction

for $N=1$
\begin{equation}
\Gamma _\alpha (z_1)\left( 1\otimes 1\right) =\exp \sum_{j\geq 1}\frac{z_1^j}%
j\alpha (-j)\exp -\sum_{j\geq 1}\frac{z_1^{-j}}j\alpha (j)e^\alpha
z_1^{\alpha \left( 0\right) }c_\alpha \quad ;  \label{25}
\end{equation}
we have the following relations: 
\begin{eqnarray}
c_\alpha \left( 1\otimes 1\right) &=&\varepsilon (\alpha ,0)1\otimes
1,\qquad \mbox{consider the convention: }\varepsilon (\alpha ,0)\equiv 1  \nonumber \\
&=&1\otimes 1 \label{26}
\end{eqnarray}
\begin{eqnarray}
z_1^{\alpha \left( 0\right) }\left( 1\otimes 1\right) &=&1\otimes \left(
e^{\ln z_1\alpha \left( 0\right) }\right) 1  \nonumber \\
&=&1\otimes 1.  \label{27}
\end{eqnarray}
where we used 
\begin{equation}
e^\alpha \left( 1\otimes 1\right) \equiv e^\alpha \left( 1\otimes e^\alpha
\right) =1\otimes e^\alpha  \label{28}
\end{equation}

From (\ref{4}) we have 
\[
u(n)\left( v^{\left( 0\right) }\otimes e^\alpha \right) \equiv u(n)\left(
1\otimes e^\alpha \right) =n\delta _{n,0}\left( u\mid v\right) \otimes
e^\alpha +\delta _{n,0}\left( \alpha \mid u\right) v^{\left( 0\right)
}\otimes e^\alpha 
\]
then for $(n>0)$ 
\begin{equation}
u(n)\left( 1\otimes 1\right) =0.\qquad  \label{29}
\end{equation}

Therefore 
\begin{eqnarray}
\exp -\sum_{j\geq 1}\frac{z_1^{-j}}j\alpha (j)\left( 1\otimes e^\alpha
\right) &=&\left[ \left( 1-\sum_{j\geq 1}\frac{z_1^{-j}}j\alpha
(j)+...\right) 1\right] \otimes e^\alpha  \nonumber \\
&=&1\otimes e^\alpha .  \label{30}
\end{eqnarray}

Besides we can write 
\[
\exp \sum_{j\geq 1}\frac{z_1^j}j\alpha (-j)\left( 1\otimes e^\alpha \right)
=\sum_{j\geq 1}\frac{z_1^j}j\alpha (-j)\otimes e^\alpha , 
\]
thus 
\begin{equation}
\Gamma _{\alpha _1}(z_1)\left( 1\otimes 1\right) =\left( \exp \sum_{j\geq 1}%
\frac{z_1^j}j\alpha _1(-j)\right) \otimes e^{\alpha _1},  \label{31}
\end{equation}
which is equal to the equation (\ref{24}) for $N=1.$

Now, let us assume that (\ref{24}) is true for a given $N$, and we will
write (\ref{24}) as 
\[
\Gamma _{\alpha _N}(z_N)\Gamma _{\alpha _{N-1}}(z_{N-1})...\Gamma _{\alpha
1}(z_1)\left( 1\otimes 1\right) =\prod_{1\leq i<j\leq N}\varepsilon (\alpha
_i,\alpha _j)\left( z_i-z_j\right) ^{\left( \alpha _i\mid \alpha _j\right)
}\cdot 
\]
\begin{equation}
\left( \prod_{i=1}^N\exp \sum_{j\in {\bf N}}\frac{z_i^j}j\alpha
_i(-j)\right) \exp \left( \sum_{i=1}^N\alpha _i\right) \left( 1\otimes
1\right) .  \label{32}
\end{equation}

Multiplying (\ref{32}) by $\Gamma _{\alpha _{N+1}}(z_{N+1})$ and using the
relations (\ref{20}), (\ref{21}) and (\ref{25})-(\ref{30}) many times, we
can write
\[
\Gamma _{\alpha _{N+1}}(z_{N+1})\Gamma _{\alpha _N}(z_N)...\Gamma _{\alpha
1}(z_1)\left( 1\otimes 1\right) =\prod_{1\leq i<j\leq N}\varepsilon (\alpha
_i,\alpha _j)\left( z_i-z_j\right) ^{\left( \alpha _i\mid \alpha _j\right)
}\cdot 
\]

\[
\prod_{1\leq k<N+1}\varepsilon (\alpha _k,\alpha _{N+1})\left(
z_k-z_{N+1}\right) ^{\left( \alpha _k\mid \alpha _{N+1}\right) }\cdot \left(
\exp \sum_{j\in {\bf N}}\frac{z_{N+!}^j}j\alpha _{N+1}(-j)\right) e^{\alpha
_{N+1}}\cdot 
\]
\[
\left( \prod_{i=1}^N\exp \sum_{j\in {\bf N}}\frac{z_i^j}j\alpha
_i(-j)\right) \exp \left( \sum_{i=1}^N\alpha _i\right) \left( 1\otimes
1\right) 
\]
\begin{eqnarray*}
&=&\prod_{1\leq i<j\leq N+1}\varepsilon (\alpha _i,\alpha _j)\left(
z_i-z_j\right) ^{\left( \alpha _i\mid \alpha _j\right) }\left(
\prod_{i=1}^{N+1}\exp \sum_{j\in {\bf N}}\frac{z_i^j}j\alpha _i(-j)\right)
\cdot \\
&&\exp \left( \sum_{i=1}^{N+1}\alpha _i\right) \left( 1\otimes 1\right) \\
&=&\prod_{1\leq i<j\leq N+1}\varepsilon (\alpha _i,\alpha _j)\left(
z_i-z_j\right) ^{\left( \alpha _i\mid \alpha _j\right) }\left(
\prod_{i=1}^{N+1}\exp \sum_{j\in {\bf N}}\frac{z_i^j}j\alpha _i(-j)\right)
\otimes \exp \left( \sum_{i=1}^{N+1}\alpha _i\right) ,
\end{eqnarray*}
which is exactly the expression (\ref{24}) written for $N+1$. This way the
equation (\ref{24}) is proved.

Moreover, making the correspondence

\[
\left| \lambda _o\right\rangle \longleftrightarrow 1\otimes 1, 
\]
where $\left| \lambda _o\right\rangle $ denotes a highest weight state of a
fundamental representation, we can write
\[
\left\langle \lambda _o\right| \Gamma _{\alpha _N}(z_N)\Gamma _{\alpha
_{N-1}}(z_{N-1})...\Gamma _{\alpha _1}(z_1)\left| \lambda _o\right\rangle = 
\]

\begin{equation}
\left\{ 
\begin{array}{c}
0,\qquad \qquad \qquad \mbox{ \qquad \qquad \qquad \qquad \qquad if }%
\sum_{i=1}^N\alpha _i\neq 0 \\ 
\prod_{1\leq i<j\leq N}\varepsilon (\alpha _i,\alpha _j)\left(
z_i-z_j\right) ^{\left( \alpha _i\mid \alpha _j\right) }\qquad \quad \mbox{%
if }\sum_{i=1}^N\alpha _i=0
\end{array}
\right.  \label{33}
\end{equation}
where we have used the fact that
\begin{eqnarray*}
\left\langle \lambda _o\right| \exp \sum_{j\in {\bf N}}\frac{z_i^j}j\alpha
_i(-j) &=&\left\langle \lambda _o\right| \left( 1+\sum_{j\in {\bf N}}\frac{%
z_i^j}j\alpha _i(-j)+...\right) \\
&=&\left\langle \lambda _o\right| ,
\end{eqnarray*}

and 
\[
\left\langle \lambda _o\right| \exp \left( \sum_{i=1}^N\alpha _i\right)
=\left\{ 
\begin{array}{c}
0,\mbox{\qquad \qquad \qquad \qquad if }\sum_{i=1}^N\alpha _i\neq 0 \\ 
\left\langle \lambda _o\right| ,\mbox{\qquad \qquad \qquad \quad if }%
\sum_{i=1}^N\alpha _i=0
\end{array}
\right. 
\]

\section{{\bf Matrix elements using a vertex operator representation of the
algebra }$\widehat{sl}_{r+1}$}

\label{appd}
Consider the correspondence 

\[
F_i\longrightarrow \nu _i\Gamma _{-\beta _i}\left( \nu _i\right) , 
\]

\[
G_i\longrightarrow \rho _i\Gamma _{\beta _i}\left( \rho _i\right) .\qquad 
\]
then we can write 
\[
C_{_{l_1...i_{l_1},}\quad j_{k_1}...j_{k_n}}=\left\langle \lambda _o\right|
F_{i_{l_1}}...F_{i_{l_n}}G_{j_{k_1}}...G_{j_{k_n}}\left| \lambda
_o\right\rangle 
\]
\[
=\nu _{i_{l_1}}...\nu _{i_{l_n}}\rho _{j_{k_1}}...\rho
_{j_{k_n}}\left\langle \lambda _o\right| \Gamma _{-\beta _{i_{l_1}}}\left(
\nu _{i_{l_1}}\right) ...\Gamma _{-\beta _{i_{l_n}}}\left( \nu
_{i_{l_n}}\right) \Gamma _{\beta _{j_{k_1}}}\left( \rho _{j_{k_1}}\right)
...\Gamma _{\beta _{j_{k_n}}}\left( \rho j_{k_n}\right) \left| \lambda
_o\right\rangle 
\]
\begin{eqnarray*}
\quad \quad &=&{\bf \{\delta }_{\overrightarrow{0},\beta _{j_{k_1}}+\cdot
\cdot \cdot \beta _{j_{k_n}}-\beta _{i_{l_1}}-\cdot \cdot \cdot \beta
_{i_{l_n}}}\nu _{i_{l_1}}...\nu _{i_{l_n}}\rho _{j_{k_1}}...\rho
_{j_{k_n}}\cdot \\
&&\prod_{1\leq a<b\leq n}\epsilon \left( \beta _{j_{k_a}},\beta
_{j_{k_b}}\right) \epsilon \left( -\beta _{i_{l_a}},-\beta _{i_{l_b}}\right)
\left( \rho _{j_{k_a}}-\rho _{j_{k_b}}\right) ^{\left( \beta _{j_{k_a}}\mid
\beta _{j_{k_b}}\right) }\cdot
\end{eqnarray*}
\begin{equation}
\left( \nu _{i_{l_a}}-\nu _{i_{l_b}}\right) ^{\left( \beta _{i_{l_a}}\mid
\beta _{i_{l_b}}\right) }{\bf \}}/{\bf \{}\prod_{1\leq a<b\leq n}\epsilon
\left( -\beta _{i_{l_a}},\beta _{j_{k_b}}\right) \left( \nu _{i_{l_a}}-\rho
_{j_{k_b}}\right) ^{\left( \beta _{i_{l_a}}\mid \beta _{j_{k_b}}\right) }%
{\bf \},}  \label{34}
\end{equation}
where we have used equation (\ref{33})$.$ 
\begin{eqnarray*}
C_{ii_{l_1}...i_{l_n},j_{k_1}...j_{k_{n+1}}}^{+} &=&\left\langle \lambda
_o\right| E_{-\beta _i}^{\left( 1\right)
}F_{i_{l_1}}...F_{i_{l_n}}G_{j_{k_1}}...G_{j_{k_{n+1}}}\left| \lambda
_o\right\rangle \\
&=&\frac 1{2\pi i}\oint d\nu .\nu \nu _{i_{l_1}}...\nu _{i_{l_n}}\rho
_{j_{k_1}}...\rho _{j_{k_{n+1}}}\cdot
\end{eqnarray*}
\[
\left\langle \lambda _o\right| \Gamma _{-\beta _i}\left( \nu \right) \Gamma
_{-\beta _{i_{l_1}}}\left( \nu _{i_{l_1}}\right) ...\Gamma _{-\beta
_{i_{l_n}}}\left( \nu _{i_{l_n}}\right) \Gamma _{\beta _{j_{k_1}}}\left(
\rho _{j_{k_1}}\right) ...\Gamma _{\beta _{j_{k_{n+1}}}}\left( \rho
j_{k_{n+1}}\right) \left| \lambda _o\right\rangle 
\]
\begin{eqnarray*}
&=&\frac 1{2\pi i}\oint d\nu .\nu \nu _{i_{l_1}}...\nu _{i_{l_n}}\rho
_{j_{k_1}}...\rho _{j_{k_{n+1}}}{\bf \delta }_{\overrightarrow{0},\beta
_{j_{k_1}}+\cdot \cdot \cdot \beta _{j_{k_{n+1}}}-\beta _i-\beta
_{i_{l_1}}-\cdot \cdot \cdot \beta _{i_{l_n}}}\cdot \\
&&{\bf \{}\prod_{0<a\leq n}\epsilon \left( -\beta _i,-\beta
_{i_{l_a}}\right) \left( \nu -\nu _{i_{l_a}}\right) ^{\left( \beta _i\mid
\beta _{i_{l_a}}\right) }\prod_{0<b\leq n+1}\epsilon \left( -\beta _i,\beta
_{j_{k_b}}\right) \left( \nu -\rho _{j_{k_b}}\right) ^{-\left( \beta _i\mid
\beta _{j_{k_b}}\right) }\cdot \\
&&\prod_{0\leq a<b\leq n}\epsilon \left( -\beta _{i_{l_a}},-\beta
_{i_{l_b}}\right) \left( \nu _{i_{l_a}}-\nu _{i_{l_b}}\right) ^{\left( \beta
_{i_{l_a}}\mid \beta _{i_{l_b}}\right) }\prod_{0\leq a<b\leq n+1}\epsilon
\left( \beta _{j_{k_a}},\beta _{j_{k_b}}\right) \cdot \\
&&
\end{eqnarray*}
\begin{equation}
\left( \rho _{j_{k_a}}-\rho _{j_{k_b}}\right) ^{-\left( \beta _{j_{k_a}}\mid
\beta _{j_{k_b}}\right) }{\bf \}}/\{\prod_{ 0\leq a\leq b\leq n  ,\, a\neq
n+1} \epsilon \left( -\beta _{i_{l_a}},\beta _{j_{k_b}}\right) \left(
\nu _{i_{l_a}}-\rho _{j_{k_b}}\right) ^{\left( \beta _{i_{l_a}}\mid \beta
_{j_{k_b}}\right) }\}.  \label{35}
\end{equation}
In the last relation we have written the following contour integration: 
\begin{equation}
I_1=\frac 1{2\pi i}\oint d\nu .\nu \frac{\prod_{0<a\leq n}\epsilon \left(
-\beta _i,-\beta _{i_{l_a}}\right) \left( \nu -\nu _{i_{l_a}}\right)
^{\left( \beta _i\mid \beta _{i_{l_a}}\right) }}{\prod_{0<b\leq n+1}\epsilon
\left( -\beta _i,\beta _{j_{k_b}}\right) \left( \nu -\rho _{j_{k_b}}\right)
^{\left( \beta _i\mid \beta _{j_{k_b}}\right) }}\cdot  \label{36}
\end{equation}

Likewise we have 
\begin{eqnarray*}
C_{_{l_1}...i_{l_{n+1}},ij_{k_1}...j_{k_n}}^{-} &=&\left\langle \lambda
_o\right| E_{\beta _i}^{\left( 1\right)
}F_{i_{l_1}}...F_{i_{l_{n+1}}}G_{j_{k_1}}...G_{j_{k_n}}\left| \lambda
_o\right\rangle \\
&=&\frac 1{2\pi i}\oint d\rho .\rho \nu _{i_{l_1}}...\nu _{i_{l_{n+1}}}\rho
_{j_{k_1}}...\rho _{j_{k_n}}\cdot
\end{eqnarray*}
\[
\left\langle \lambda _o\right| \Gamma _{\beta _i}\left( \rho \right) \Gamma
_{-\beta _{i_{l_1}}}\left( \nu _{i_{l_1}}\right) ...\Gamma _{-\beta
_{i_{l_{n+1}}}}\left( \nu _{i_{l_{n+1}}}\right) \Gamma _{\beta
_{j_{k_1}}}\left( \rho _{j_{k_1}}\right) ...\Gamma _{\beta _{j_{k_n}}}\left(
\rho j_{k_n}\right) \left| \lambda _o\right\rangle 
\]
\begin{eqnarray*}
&=&\frac 1{2\pi i}\oint d\rho .\rho \nu _{i_{l_1}}...\nu _{i_{l_{n+1}}}\rho
_{j_{k_1}}...\rho _{j_{k_n}}{\bf \delta }_{\overrightarrow{0},\beta _i+\beta
_{j_{k_1}}+\cdot \cdot \cdot \beta _{j_{k_n}}-\beta _{i_{l_1}}-\cdot \cdot
\cdot \beta _{i_{l_{n+1}}}}\cdot \\
&&{\bf \{}\prod_{0<a\leq n}\epsilon \left( \beta _i,\beta _{j_{k_a}}\right)
\left( \rho -\rho _{j_{k_a}}\right) ^{\left( \beta _i\mid \beta
_{j_{k_a}}\right) }\prod_{0<b\leq n+1}\epsilon \left( \beta _i,\beta
_{i_{l_b}}\right) \left( \rho -\nu _{i_{l_b}}\right) ^{-\left( \beta _i\mid
\beta _{i_{l_b}}\right) }\cdot \\
&&\prod_{0\leq a<b\leq n}\epsilon \left( \beta _{j_{k_a}},\beta
_{j_{k_b}}\right) \left( \rho _{j_{k_a}}-\rho _{j_{k_b}}\right) ^{\left(
\beta _{j_{k_a}}\mid \beta _{j_{k_b}}\right) }\prod_{0\leq a<b\leq
n+1}\epsilon \left( -\beta _{i_{l_a}},-\beta _{i_{l_b}}\right) \cdot
\end{eqnarray*}
\begin{equation}
\left( \nu _{i_{l_a}}-\nu _{i_{l_b}}\right) ^{-\left( \beta _{i_{l_a}}\mid
\beta _{i_{l_b}}\right) }{\bf \}}/\{\prod_{ 0\leq a\leq b\leq n  ,\, b\neq
n+1  } \epsilon \left( -\beta _{i_{l_a}},\beta _{j_{k_b}}\right) \left(
\nu _{i_{l_a}}-\rho _{j_{k_b}}\right) ^{\left( \beta _{i_{l_a}}\mid \beta
_{j_{k_b}}\right) }\}.  \label{37}
\end{equation}
This time the relevant contour integration becomes 
\begin{equation}
I_2=\frac 1{2\pi i}\oint d\rho .\rho \frac{\prod_{0<a\leq n}\epsilon \left(
\beta _i,-\beta _{j_{k_a}}\right) \left( \rho -\rho _{j_{k_a}}\right)
^{\left( \beta _i\mid \beta _{j_{k_a}}\right) }}{\prod_{0<b\leq n+1}\epsilon
\left( \beta _i,\beta _{i_{l_b}}\right) \left( \rho -\nu _{i_{l_b}}\right)
^{\left( \beta _i\mid \beta _{i_{l_b}}\right) }}\cdot  \label{38}
\end{equation}

\section{{\bf Matrix elements using the homogeneous vertex operator
representation of the Kac-Moody algebra }$\widehat{sl}_2$}

\label{appe}
In the case of the affine algebra $\widehat{sl}_2$, we have the generators

\[
\alpha =\left( 
\begin{array}{cc}
1 & 0 \\ 
0 & -1
\end{array}
\right) ,\quad e=\left( 
\begin{array}{cc}
0 & 1 \\ 
0 & 0
\end{array}
\right) ,\quad f=\left( 
\begin{array}{cc}
0 & 0 \\ 
1 & 0
\end{array}
\right) 
\]
and let us choose the following dual bases of $\widehat{sl}_2$%
\[
\left\{ t^n\alpha ,t^ne,t^nf,C,D\right\} ,\quad \mbox{and \quad }\left\{
\frac 12t^{-n}\alpha ,t^{-n}f,t^{-n}e,D,C\right\} . 
\]
We have 
\begin{eqnarray*}
Q &=&{\bf \ Z}\alpha ,\quad \left( \alpha \mid \alpha \right) =2,\quad
\varepsilon (\alpha ,\alpha )=\varepsilon (-\alpha ,-\alpha )=-1,\quad
\varepsilon (\alpha ,\alpha )=\varepsilon (-\alpha ,-\alpha )=1\, \\
&&\mbox{(we are using the ``gauge fixing'' of \ct{goddard, frenkel} for }\varepsilon (\alpha
,\beta)).
\end{eqnarray*}

Considering $q=e^\alpha ,$ we identify ${\bf C}\left[ Q\right] $ with ${\bf C}%
\left[ q,q^{-1}\right] .$ Thus the homogeneous vertex operator construction can be described as follows 
\[
L\left( \lambda _o\right) ={\bf C}\left[ x_1,x_2,...;q,q^{-1}\right] ; 
\]
\begin{eqnarray*}
\alpha ^{\left( n\right) } &\mapsto &2\frac \partial {\partial x_n}\mbox{
\quad and \quad }\alpha ^{\left( -n\right) }\mapsto nx_n\mbox{ \quad for
\quad }n>0,\qquad \alpha ^{\left( 0\right) }\mapsto 2q\frac \partial
{\partial q}; \\
C &\mapsto &1,\qquad D\mapsto -\left( q\frac \partial {\partial q}\right)
^2-\sum_{n\geq 1}nx_n\frac \partial {\partial x_n};
\end{eqnarray*}
\[
E(z):=\sum_{n\in {\bf Z}}E_{\pm }^{\left( n\right) }z^{-n-1}\mapsto \Gamma
_{\pm }(z) 
\]
where 
\begin{equation}
\Gamma _{\pm }(z)=\exp \left( \pm \sum_{j\geq 1}z^jx_j\right) \exp \left(
\mp 2\sum_{j\geq 1}\frac{z^{-j}}j\frac \partial {\partial x_j}\right) q^{\pm
1}z^{\pm 2q\frac \partial {\partial q}}c_{\pm \alpha },  \label{39}
\end{equation}
(note that $z^{\pm 2q\frac \partial {\partial q}}\left( q^n\right) =z^{\pm
2n}q^n$).

Then, we can make use of the Eqn.(\ref{33}) to compute the matrix elements,
 viz.,

\br
\left\langle \lambda _o\right| \Gamma _{\alpha _N}(z_N)\Gamma _{\alpha
_{N-1}}(z_{N-1})...\Gamma _{\alpha _1}(z_1)\left| \lambda _o\right\rangle = 
\er

\begin{equation}
\left\{ 
\begin{array}{c}
0,\qquad \qquad \qquad \mbox{ \qquad \qquad \qquad \qquad \qquad if }%
\sum_{i=1}^N\alpha _i\neq 0 \\ 
\prod_{1\leq i<j\leq N}\varepsilon (\alpha _i,\alpha _j)\left(
z_i-z_j\right) ^{\left( \alpha _i\mid \alpha _j\right) }\qquad \quad \mbox{%
if }\sum_{i=1}^N\alpha _i=0
\end{array}
\right.  \label{40}
\end{equation}

In this way we have two types of operators, $\Gamma _{\pm }(z)$, associated to $%
\alpha $ and -$\alpha $ respectively. We should have an even number of $%
\Gamma ^{,}s$ in Eqn. (\ref{33}) in order to have a non zero value for $%
\left\langle \lambda _o\right| \Gamma _{\alpha _N}(z_N)\Gamma _{\alpha
_{N-1}}(z_{N-1})...\Gamma _{\alpha _1}(z_1)\left| \lambda _o\right\rangle $;
thus we may choose $2N$ operators, such that $N$ operators correspond to $\alpha 
$ and the remaining $N$ of them correspond to $-\alpha .$

We provide some of the matrix components we used in the construction of {\sl one-soliton} and {\sl two-soliton} solutions of the system {\bf NLS}. Defining
\br
G_i=\sum_{n=-\infty }^{+\infty }\nu _i^{-n}E_{+}^{\left( n\right) },\qquad
F_i=\sum_{n=-\infty }^{+\infty }\nu _i^{-n}E_{-}^{\left( n\right) }; 
\er
we can make the correspondence 
\br
G_i\longrightarrow \rho _i\Gamma _{+}\left( \rho _i\right) ,\qquad
F_i\longrightarrow \nu _i\Gamma _{-}\left( \nu _i\right) . 
\er
Then
\begin{eqnarray*}
\left\langle \lambda _o\right| G_iF_j\left| \lambda _o\right\rangle
&=&\left\langle \lambda _o\right| F_jG_i\left| \lambda _o\right\rangle =\rho
_i\nu _j\left\langle \lambda _o\right| \Gamma _{+}(\rho _i)\Gamma _{-}(\nu
_j)\left| \lambda _o\right\rangle \\
&=&\rho _i\nu _j\frac{\varepsilon (+,-)}{\left( \rho _i-\nu _j\right) ^2} \\
&=&\frac{\rho _i\nu _j}{\left( \rho _i-\nu _j\right) ^2}\cdot
\end{eqnarray*}

As a special case we compute the following expression
\begin{eqnarray*}
\left\langle \lambda _o\right| E_{-}^1G_i\left| \lambda _o\right\rangle
&=&\frac 1{2\pi i}\oint dz\,z\rho _i\left\langle \lambda _o\right| \Gamma
_{-}(z)\Gamma _{+}(\rho _i)\left| \lambda _o\right\rangle \\
&=&\frac 1{2\pi i}\rho _i\oint dz.\frac z{\left( z-\rho _i\right) ^2} \\
&=&\rho _i,
\end{eqnarray*}
where we have used 
\[
E_{-}^1=\oint dz\,z\Gamma _{-}(z), 
\]
where integration is over some curve encircling the origin.

The same method can be used to compute the following matrix element
\begin{eqnarray*}
\left\langle \lambda _o\right| E_{-}^1G_1F_2G_2\left| \lambda
_o\right\rangle &=&\frac 1{2\pi i}\oint dz\,z\,\rho _{1}\,\nu _{2}\,\rho
_2\left\langle \lambda _o\right| \Gamma _{-}(z)\Gamma _{+}(\rho _1)\Gamma
_{-}(\nu _2)\Gamma _{+}(\rho _2)\left| \lambda _o\right\rangle \\
&=&\frac 1{2\pi i}\oint dz\,z\,\rho _{1}\,\nu _{2}\,\rho _2\varepsilon
(+,-)\varepsilon (+,+)\varepsilon (+,-)\varepsilon (-,+)\varepsilon
(-,-)\varepsilon (+,-) \\
&&\frac{\left( \rho _2-\rho _1\right) ^2\left( \nu _2-z\right) ^2}{\left(
\rho _2-\nu _2\right) ^2\left( \nu _2-\rho _1\right) ^2\left( \rho
_2-z\right) ^2\left( \rho _1-z\right) ^2} \\
&=&\rho _{1}\,\nu _{2}\,\rho _2\frac{\left( \rho _2-\rho _1\right) ^2}{\left(
\rho _2-\nu _2\right) ^2\left( \nu _2-\rho _1\right) ^2}\cdot
\end{eqnarray*}

The remaining matrix elements can be computed in the same way. We give some
of them 
\begin{eqnarray*}
\left\langle \lambda _o\right| E_{-}^1G_i\left| \lambda _o\right\rangle
&=&\frac 1{2\pi i}\oint dz.z\rho _i\left\langle \lambda _o\right| \Gamma
_{-}(z)\Gamma _{+}(\rho _i)\left| \lambda _o\right\rangle \qquad \\
&=&\frac 1{2\pi i}\rho _i\oint dz\,\frac z{\left( z-\rho _i\right) ^2} \\
&=&\rho _i\,\quad \qquad \qquad \qquad
\end{eqnarray*}

\br
\left\langle \lambda _o\right| F_iG_iF_jG_j\left| \lambda _o\right\rangle =%
\frac{\rho _{i}\,\nu _{j}\,\rho _i\rho _j\left( \rho _j-\rho _i\right)
^2\left( \nu _j-\nu _i\right) ^2}{\left( \rho _j-\nu _j\right) ^2\left( \rho
_j-\nu _i\right) ^2\left( \rho _i-\nu _i\right) ^2\left( \rho _i-\nu
_j\right) ^2}, 
\er

\br
\left\langle \lambda _o\right| E_{-}^1G_iF_jG_j\left| \lambda
_o\right\rangle =\frac{\rho _{i}\,\nu _{j}\,\rho _j\left( \rho _j-\rho
_i\right) ^2}{\left( \rho _j-\nu _j\right) ^2\left( \nu _j-\rho _i\right) ^2}%
,\qquad \qquad \qquad \qquad \quad 
\er

\br
\left\langle \lambda _o\right| E_{+}^1F_i\left| \lambda _o\right\rangle =\nu
_i,\qquad \qquad \qquad \qquad \qquad \qquad \qquad \qquad 
\er

\br
\left\langle \lambda _o\right| E_{+}^1F_iG_iF_j\left| \lambda
_o\right\rangle =\frac{\nu _{i}\,\nu _{j}\,\rho _i\left( \nu _j-\nu _i\right)
^2}{\left( \rho _i-\nu _j\right) ^2\left( \nu _i-\rho _i\right) ^2}\cdot
\qquad \qquad \qquad \qquad \quad 
\er

These results suggest that a general matrix element could be expressed in terms of Vandermonde-like determinants. In fact, recently in \ct{meinel} there was derived a natural relationship with the {\sl Vandermonde-like determinants}. The resulting framework in our case may be well-suited to achieve a compactness and transparency in $N$-soliton formulas.

\section{{\bf The contour integration of the matrix elements in the case of }%
$\widehat{sl}_2$}
\label{appf}
Let us show that the integral 
\begin{equation}
I_n=\frac 1{2\pi i}\oint dz.z\frac{\left( z-\nu _1\right) ^2...\left( z-\nu
_{n-1}\right) ^2}{\left( z-\rho _1\right) ^2...\left( z-\rho _n\right) ^2}%
,\qquad n\geq 1  \label{41}
\end{equation}
is equal to unity. We will prove by induction.

For $n=1:$

\begin{eqnarray*}
I_1 &=&\frac 1{2\pi i}\oint dz\frac z{\left( z-\rho _1\right) ^2} \\
&=&1.
\end{eqnarray*}
Assuming that $I_n=1$ for some $n>1$ let us determine the form of $I_{n+1}$.

The expression for $I_{n+1}$ can be written as 
\begin{equation}
I_{n+1}=I_n+C_n,  \label{42}
\end{equation}
where 
\[
C_n=\frac 1{2\pi i}\frac \partial {\partial \rho _1}\cdot \cdot \cdot \frac
\partial {\partial \rho _{n+1}}\oint dz.z\frac{\left( z-\nu _1\right)
^2\cdot \cdot \cdot \left( z-\nu _{n-1}\right) ^2}{\left( z-\rho _1\right)
\cdot \cdot \cdot \left( z-\rho _{n+1}\right) }\left( \rho _{n+1}-\nu
_n\right) ^2 
\]
\begin{eqnarray*}
&=&\frac 1{2\pi i}{\bf \{}\frac \partial {\partial \rho _1}\cdot \cdot \cdot
\frac \partial {\partial \rho _{n+1}}\oint dz\left( \rho _2-\nu _1\right)
^2\left( \rho _3-\nu _2\right) ^2\cdot \cdot \cdot \left( \rho _{n+1}-\nu
_n\right) ^2\cdot \\
&&\frac z{\left( z-\rho _1\right) \cdot \cdot \cdot \left( z-\rho
_{n+1}\right) }+ \\
&&{\bf [}\frac \partial {\partial \rho _1}\frac \partial {\partial \rho
_3}\cdot \cdot \cdot \frac \partial {\partial \rho _{n+1}}\oint dz\left(
\rho _3-\nu _2\right) ^2\cdot \cdot \cdot \left( \rho _{n+1}-\nu _n\right)
^2\cdot \\
&&\frac z{\left( z-\rho _1\right) \left( z-\rho _3\right) \cdot \cdot \cdot
\left( z-\rho _{n+1}\right) }+ \\
&&\frac \partial {\partial \rho _1}\frac \partial {\partial \rho _2}\frac
\partial {\partial \rho _4}\cdot \cdot \cdot \frac \partial {\partial \rho
_{n+1}}\oint dz\left( \rho _2-\nu _1\right) ^2\left( \rho _4-\nu _3\right)
^2\cdot \cdot \cdot \left( \rho _{n+1}-\nu _n\right) ^2\cdot \\
&&\frac z{\left( z-\rho _1\right) \left( z-\rho _2\right) \left( z-\rho
_4\right) \cdot \cdot \cdot \left( z-\rho _{n+1}\right) }+\cdot \cdot \cdot +
\\
&&\frac \partial {\partial \rho _1}\ \cdot \cdot \cdot \frac \partial
{\partial \rho _{n-1}}\frac \partial {\partial \rho _{n+1}}\oint dz\left(
\rho _2-\nu _1\right) ^2\left( \rho _3-\nu _2\right) ^2\cdot \cdot \cdot
\left( \rho _{n-1}-\nu _{n-2}\right) ^2\left( \rho _{n+1}-\nu _n\right)
^2\cdot \\
&&\frac z{\left( z-\rho _1\right) \cdot \cdot \cdot \left( z-\rho
_{n-1}\right) \left( z-\rho _{n+1}\right) }{\bf ]}+{\bf [}\frac \partial
{\partial \rho _1}\frac \partial {\partial \rho _4}\cdot \cdot \cdot \frac
\partial {\partial \rho _{n+1}}\cdot \\
&&\oint dz\left( \rho _4-\nu _3\right) ^2\cdot \cdot \cdot \left( \rho
_{n+1}-\nu _n\right) ^2\cdot \\
&&\frac z{\left( z-\rho _1\right) \left( z-\rho _4\right) \cdot \cdot \cdot
\left( z-\rho _{n+1}\right) }+\frac \partial {\partial \rho _1}\frac
\partial {\partial \rho _3}\frac \partial {\partial \rho _5}\cdot \cdot
\cdot \frac \partial {\partial \rho _{n+1}}\cdot \\
&&\oint dz\left( \rho _3-\nu _2\right) ^2\left( \rho _5-\nu _4\right)
^2\cdot \cdot \cdot \left( \rho _{n+1}-\nu _n\right) ^2\frac z{\left( z-\rho
_1\right) \left( z-\rho _3\right) \left( z-\rho _5\right) \cdot \cdot \cdot
\left( z-\rho _{n+1}\right) }+ \\
&&\frac \partial {\partial \rho _1}\frac \partial {\partial \rho _3}\cdot
\cdot \cdot \frac \partial {\partial \rho _{n-1}}\frac \partial {\partial
\rho _{n+1}}\oint dz\left( \rho _3-\nu _2\right) ^2\cdot \cdot \cdot \left(
\rho _{n-1}-\nu _{n-2}\right) ^2\left( \rho _{n+1}-\nu _n\right) ^2\cdot
\end{eqnarray*}
\[
\cdot \cdot \cdot + 
\]
\begin{eqnarray*}
&&\frac z{\left( z-\rho _1\right) \left( z-\rho _3\right) \cdot \cdot \cdot
\left( z-\rho _{n-1}\right) \left( z-\rho _{n+1}\right) }+\frac \partial
{\partial \rho _1}\frac \partial {\partial \rho _2}\frac \partial {\partial
\rho _5}\cdot \cdot \cdot \frac \partial {\partial \rho _{n+1}}\cdot \\
&&\oint dz\left( \rho _2-\nu _1\right) ^2\left( \rho _5-\nu _4\right)
^2\cdot \cdot \cdot \left( \rho _{n+1}-\nu _n\right) ^2\frac z{\left( z-\rho
_1\right) \left( z-\rho _2\right) \left( z-\rho _5\right) \cdot \cdot \cdot
\left( z-\rho _{n+1}\right) }+ \\
&&\cdot \cdot \cdot +\frac \partial {\partial \rho _1}\frac \partial
{\partial \rho _2}\frac \partial {\partial \rho _4}\cdot \cdot \cdot \frac
\partial {\partial \rho _{n-1}}\frac \partial {\partial \rho _{n+1}}\oint
dz\left( \rho _2-\nu _1\right) ^2\left( \rho _4-\nu _3\right) ^2\cdot \cdot
\cdot \left( \rho _{n-1}-\nu _{n-2}\right) ^2\cdot \\
&&\left( \rho _{n+1}-\nu _n\right) ^2\frac z{\left( z-\rho _1\right) \left(
z-\rho _2\right) \left( z-\rho _4\right) \cdot \cdot \cdot \left( z-\rho
_{n-1}\right) \left( z-\rho _{n+1}\right) }+\cdot \cdot \cdot + \\
&&\frac \partial {\partial \rho _1}\ \cdot \cdot \cdot \frac \partial
{\partial \rho _{n-2}}\frac \partial {\partial \rho _{n+1}}\oint dz\left(
\rho _2-\nu _1\right) ^2\cdot \cdot \cdot \left( \rho _{n-2}-\nu
_{n-3}\right) ^2\left( \rho _{n+1}-\nu _n\right) ^2\cdot \\
&&\frac z{\left( z-\rho _1\right) \cdot \cdot \cdot \left( z-\rho
_{n-2}\right) \left( z-\rho _{n+1}\right) }{\bf ]}+{\bf [}\frac \partial
{\partial \rho _1}\frac \partial {\partial \rho _5}\cdot \cdot \cdot \frac
\partial {\partial \rho _{n+1}}\oint dz\left( \rho _5-\nu _4\right) ^2\cdot
\cdot \cdot \\
&&\left( \rho _{n+1}-\nu _n\right) ^2\frac z{\left( z-\rho _1\right) \left(
z-\rho _5\right) \cdot \cdot \cdot \left( z-\rho _{n+1}\right) }+\frac
\partial {\partial \rho _1}\frac \partial {\partial \rho _4}\frac \partial
{\partial \rho _5}\cdot \cdot \cdot \frac \partial {\partial \rho
_{n-1}}\frac \partial {\partial \rho _{n+1}}\cdot \\
&&\oint dz\left( \rho _4-\nu _3\right) ^2\left( \rho _5-\nu _4\right)
^2\cdot \cdot \cdot \left( \rho _{n-1}-\nu _{n-2}\right) ^2\left( \rho
_{n+1}-\nu _n\right) ^2\cdot \\
&&\frac z{\left( z-\rho _1\right) \left( z-\rho _4\right) \left( z-\rho
_5\right) \cdot \cdot \cdot \left( z-\rho _{n-1}\right) \left( z-\rho
_{n+1}\right) }+\cdot \cdot \cdot +
\end{eqnarray*}
\begin{eqnarray}
&&\frac \partial {\partial \rho _1}\ \cdot \cdot \cdot \frac \partial
{\partial \rho _{n-3}}\frac \partial {\partial \rho _{n+1}}\oint dz\left(
\rho _2-\nu _1\right) ^2\cdot \cdot \cdot \left( \rho _{n-3}-\nu
_{n-4}\right) ^2\left( \rho _{n+1}-\nu _n\right) ^2\cdot  \nonumber \\
&&\frac z{\left( z-\rho _1\right) \cdot \cdot \cdot \left( z-\rho
_{n-3}\right) \left( z-\rho _{n+1}\right) }{\bf ]+\cdot \cdot \cdot +} 
\nonumber \\
&&\frac \partial {\partial \rho _1}\frac \partial {\partial \rho
_{n+1}}\oint dz\left( \rho _{n+1}-\nu _n\right) ^2\frac z{\left( z-\rho
_1\right) \left( z-\rho _{n+1}\right) }{\bf \}\cdot }  \label{43}
\end{eqnarray}
It is easy to show the following:

{\bf i)} 
\begin{equation}
\frac 1{2\pi i}\oint dz\frac z{\left( z-\rho _1\right) \left( z-\rho
_2\right) }=1,  \label{44}
\end{equation}
and
{\bf ii)} 
\begin{equation}
\oint dz\frac z{\left( z-\rho _1\right) \cdot \cdot \cdot \left( z-\rho
_n\right) }=0\qquad \mbox{for\qquad }n>2\cdot  \label{45}
\end{equation}
writing 
\begin{equation}
\frac 1{\left( z-\rho _1\right) \cdot \cdot \cdot \left( z-\rho _n\right)
}=\frac 1{\det \Delta }\left[ \frac{A_{n,1}}{z-\rho _1}+\frac{A_{n,2}}{%
z-\rho _2}+\cdot \cdot \cdot +\frac{A_{n,n}}{z-\rho _n}\right] ,  \label{46}
\end{equation}
with 
\br
\nonu
&&\Delta =\\
\nonu
&&\left( 
\begin{array}{cccc}
1 & 1 & ...& 1\\ 
-\sum_{i=1,\, i\neq 1}^n\rho _i  & - \sum_{ i=1  ,\, i\neq 2} ^n\rho _i & \cdot \cdot \cdot  & 
- \sum_{ i=1 , \, i\neq n} ^n\rho _i\\
  \sum_{ 1\leq i<j\leq n  ,\, i,j\neq 1 } ^n\rho _i\rho
_j  & \sum_{ 1\leq i<j\leq n  ,\, i,j\neq 2} ^n\rho
_i\rho _j & \cdot \cdot \cdot  & \sum_{ 1\leq i<j\leq n 
,\, i,j\neq n} ^n\rho _i\rho _j  \\ 
- \sum_{ 1\leq i<j<k\leq n  ,\, i,j,k\neq 1} ^n\rho _i\rho
_j\rho _k & - \sum_{ 1\leq i<j<k\leq n  ,\, i,j,k\neq 2 
} ^n\rho _i\rho _j\rho _k & \cdot \cdot \cdot  & -
\sum_{ 1\leq i<j<k\leq n  ,\, i,j,k\neq n} ^n\rho _i\rho _j\rho_k \\ 
 \cdot  & \cdot  &  &  \cdot  \\ 
 \cdot  & \cdot  &  &  \cdot \\ 
 \cdot  & \cdot  &  &  \cdot \\ 
 (-1)^{n-1}\rho _2\rho _3...\rho _n & \left( -1\right)
^{n-1}\rho _1\rho _3\rho _4...\rho _n & \cdot \cdot \cdot  & \left(
-1\right)^{n-1}\rho _1\rho _2...\rho _{n-1}
\end{array}
\right)\\
\label{47}
\er
$A_{i,j}$ denotes the cofactor of the element $\Delta _{i,j}.$ Let us note
that, multiplying by $\sum_{i=1}^n\rho _i$ the first row and adding to the
second row of the matrix $\Delta ,$ the $\det \Delta $ and the cofators $%
A_{n,i}$ do not change$;$ then (\ref{46}), can be written as

\begin{equation}
\frac 1{\left( z-\rho _1\right) \cdot \cdot \cdot \left( z-\rho _n\right)
}=\frac 1{\det d}\left[ \frac{B_{n,1}}{z-\rho _1}+\frac{B_{n,2}}{z-\rho _2}%
+\cdot \cdot \cdot +\frac{B_{n,n}}{z-\rho _n}\right]  \label{48}
\end{equation}
where $\det \Delta =\det d$ and $A_{i,j}=B_{i,j},\,\,$ $B_{i,j}$
are the cofactors of the matrix elements $d_{i,j}.$ It is easy to realize
that the matrix $d$ has $d_{2,i}=\rho _i$ as the elements of the second row.
Then the contour integration of the relevant terms of (\ref{48}) is 
\[
\frac 1{2\pi i}\oint dz\frac z{\left( z-\rho _1\right) ...\left( z-\rho
_n\right) }=\frac 1{\det d}\left[ \rho _1B_{n,1}+\rho _2B_{n,2}+\cdot \cdot
\cdot +\rho _nB_{n,n}\right] . 
\]

We know that 
\[
\sum_{i=1}^nd_{l,j}B_{k,j}=0,\qquad l\neq k 
\]
and therefore, we can conclude 
\begin{equation}
\oint dz\frac z{\left( z-\rho _1\right) \cdot \cdot \cdot \left( z-\rho
_n\right) }=0,\qquad \mbox{for }n>2  \label{49}
\end{equation}

Thus, the last term of $C_n$ vanishes due to (\ref{44}) and the remaining
terms vanish because of (\ref{45}). Then $C_n=0$ for $n>1,$ and the Eqn.(\ref
{42}) becomes

\[
I_{n+1}=I_n,\qquad n>1, 
\]
this shows that
\[
I_n=1\qquad n\geq 1. 
\]

\end{document}